\documentclass[a4paper,fleqn,usenatbib]{mnras}

\usepackage{mathptmx}

\usepackage[T1]{fontenc}
\usepackage{ae,aecompl}

\usepackage{graphicx}	
\usepackage{amsmath}	
\usepackage{amssymb}	
\usepackage[english]{babel}
\usepackage{xspace}

\newcommand{\TotalSample}{981\xspace}   
\newcommand{\NUVrob}{245\xspace}
\newcommand{\RadioDet}{512\xspace}
\newcommand{\SNfit}{3\xspace}          
\newcommand{\WISESNnum}{552\xspace}    
\newcommand{\SubCapt}{The panel is split into: full sample (top left), non line emitters (top right), strong line emitters (bottom left) and weak line emitters (bottom right.) The (blue) squares indicate those BCGs with colour offsets in excess of $>2.5\sigma_p$ from zero, where $\sigma_p$ is the scatter of the non-line emitters.\xspace}    
\newcommand{\wb}{\textit{W}}    
\newcommand{\ib}{\textit{i}\xspace}  
\newcommand{\grb}{\textit{g}--\textit{r}\xspace}   
\newcommand{\rib}{\textit{r}--\textit{i}\xspace}     
\newcommand{\izb}{\textit{i}--\textit{z}\xspace}     
\newcommand{\wc}{\textit{W}1--\textit{W}2\xspace}    
\newcommand{\wcol}{\textit{W}2--\textit{W}3\xspace}   
\newcommand{\Mpc}{\ensuremath{\mathrm{Mpc}}}  
\newcommand{\galex}{\textit{GALEX}\xspace}  
\newcommand{\wise}{\textit{WISE}\xspace}  
\newcommand{\Lx}{$L_{\mathrm{X}}$\xspace}      

\DeclareGraphicsExtensions{.pdf, .png}     

\title[A Photometric Census of Brightest Cluster Galaxies]{A Multi-Wavelength Photometric Census of AGN and Star Formation Activity in the Brightest Cluster Galaxies of X-ray Selected Clusters}
\author[Green et al.]{
T. S. Green$^{1}$\thanks{E-mail:t.s.green@durham.ac.uk}, 
A. C. Edge$^{1}$,
J. P. Stott$^{1,3}$, 
H. Ebeling$^{2}$, 
W. S. Burgett$^{4}$,
\and K. C. Chambers$^{2}$,
P. W. Draper$^{1}$,
N. Metcalfe$^{1}$,
N. Kaiser$^{2}$,
R. J. Wainscoat$^{2}$,
\and C. Waters$^{2}$
\\
$^{1}$Centre for Extragalactic Astronomy, Durham University, South Road, Durham DH1 3LE, UK\\
$^{2}$Institute for Astronomy, University of Hawaii, 2680 Woodlawn Drive, Honolulu, HI 96822, USA\\
$^{3}$Sub-department of Astrophysics, Department of Physics, University of Oxford, Denys Wilkinson Building, Keble Road, Oxford OX1 3RH, UK \\
$^{4}$GMTO Corporation, 465 N. Halstead St., Suite 250, Pasadena, CA  91107, USA
}

\date{Accepted 2016 June 2. Received 2016 May 31; in original form 2016 January 29}
\pubyear{2016}

\begin{document}
\label{firstpage}
\pagerange{\pageref{firstpage}--\pageref{lastpage}}
\maketitle

\begin{abstract}
Despite their reputation as being ``red and dead'', the unique environment inhabited by Brightest Cluster Galaxies (BCGs) can often lead to a self-regulated feedback cycle between radiatively cooling intracluster gas and star formation and AGN activity in the BCG. However the prevalence of ``active'' BCGs, and details of the feedback involved, are still uncertain. We have performed an optical, UV and Mid-IR photometric analysis of the BCGs in \TotalSample clusters at $0.03<z<0.5$, selected from the \textit{ROSAT} All Sky Survey. Using Pan-STARRS PS1 $3\pi$, \galex and \wise survey data we look for BCGs with photometric colours which deviate from that of the bulk population of passive BCGs - indicative of AGN and/or star formation activity within the BCG. We find that whilst the majority of BCGs are consistent with being passive, at least $14\%$ of our BCGs show a significant colour offset from passivity in at least one colour index. And, where available, supplementary spectroscopy reveals the majority of these particular BCGs show strong optical emission lines. On comparing BCG ``activity'' with the X-ray luminosity of the host cluster, we find that BCGs showing a colour offset are preferentially found in the more X-ray luminous clusters, indicative of the connection between BCG ``activity'' and the intracluster medium.  
\end{abstract}


\section{Introduction}
As the most massive galaxies in the Universe, and positioned at the centre of the cluster potential, the evolution of Brightest Cluster Galaxies (BCGs) is affected by the Intracluster Medium (ICM). In turn, feedback effects from Active Galactic Nuclei (AGN) and star formation (SF) activity in the BCG affects the evolution of the ICM. A self-regulated cycle of radiative cooling and AGN heating is widely acknowledged, however, the physics of this heating mechanism are not well understood. With photometric surveys becoming ever larger, identification of active BCG candidates through photometry is crucial in our attempts to understand the important role that feedback plays in galaxy evolution.  
\\ \indent Early-type galaxies in the cores of galaxy clusters are typically referred to as ``red and dead'' galaxies - their stellar populations have been passively evolving since forming in an essentially instantaneous burst of star formation at high redshift ($z>2$) (e.g. \citealt{Stanford+98,Andreon03,Thomas+05,vanDokkum+07,Mei+09}). Located at the centre of the cluster potential however, and surrounded by dense intracluster gas, BCGs occupy a unique environment, with an enhanced likelihood to foster cool gas and star formation.
\\ \indent The intracluster medium (ICM), which outweighs the constituent galaxies by a factor of 10 in rich clusters \citep{Lin+03}, exists at typical temperatures of $T\sim10^{7-8}\,\mathrm{K}$ and, as a result, radiates away huge amounts of energy through thermal Bremsstrahlung X-ray emission (\Lx $\sim10^{43-45}\,\mathrm{erg s^{-1}}$). Models predict that an unopposed process of such radiative cooling would lead to gas cooling through intermediate temperatures, a reservoir of cold gas in the centre of clusters, and subsequent star formation. However, X-ray observations with \textit{XMM-Newton} and \textit{Chandra} of clusters with very peaked central emission, and hence cooling time-scales shorter than the age of the system - known as ``cool core'' clusters - reveal that this is not the case. The observed temperatures of the ``cool cores'' are rarely lower than 1/3 of the virial temperature, warmer than pure cooling models would predict, and the mass of cooling gas and the star formation rates are lower than predicted. As the deposition rate of cool gas in such clusters was expected to cause a flow of gas of the order of $100\mathrm{s-}1000\mathrm{s}\,M_{\sun}\mathrm{yr}^{-1}$ \citep{Fabian94}, this came to be known as the ``cooling flow'' problem. The realisation now is that a heating mechanism must exist, offsetting the cooling rate, and that a self regulated cycle of cooling and heating occurs. The primary source of this heating is thought to be AGN feedback (\citealt{Fabian12,McNamara12}.
\\ \indent The majority of cool core clusters are observed to have a central radio galaxy (\citealt{Burns+90,Hogan+15}) and heating via this AGN activity is supported by observations of co-located X-ray cavities and radio lobes around radio bright BCGs (e.g. \citealt{Boehringer+93, McNamara+00, HL+15}). Cavities, resulting from the displacement of the X-ray emitting ICM by jet driven outflowing plasma, are seen in $\geq70\%$ of cool core clusters (\citealt{Dunn+06,Hlavacek+12}). The energy necessary to create these cavities is generally sufficient to balance the energy loss through radiative cooling (\citealt{Rafferty+06,McNamara+07}). The exact nature of this heating is however, still disputed with several mechanisms proposed (see \citealt{Fabian12} and \citealt{McNamara12} for reviews). Despite the fact it is common to see radio emission from BCGs, and the evidence suggests that the AGN often play an important role over time-scales of at least 1 Gyr, BCGs which exhibit signs of an ongoing strong AGN outburst are rare because of the short AGN duty cycle. Hence, in order to constrain the prevalence of ongoing AGN in BCGs, over a wide range of AGN power, it is necessary to have a large sample of BCGs. 
\\ \indent Although the star formation rate in BCGs is lower than pure cooling models predict, an enhanced star formation is observed in the BCGs of many clusters (e.g. \citealt{Egami+06IR,Donahue+10,Hicks+10,Liu+12, Fogarty+15}), which correlates with the gas properties of the host cluster. For example, \citet{Wang+10} find that when comparing optically, and X-ray, selected BCGs, along with non-BCGs, that only X-ray selected BCGs show an enhanced star formation, particularly in cool cores, suggesting that the thermodynamic state of gas is important in BCG activity. Infrared (IR) emission in BCGs, for example, is seen to anti-correlate with the X-ray cooling time in the cluster core, both in the far-IR \citep{Rawle+12}, and the Mid-IR \citep{Odea+08}. \citet{Hoffer+12} find that below an central gas entropy of 30keV cm$^2$, BCGs are more likely to show an UV excess and enhanced Mid-IR emission. This apparent entropy threshold is in agreement with \citet{Rafferty+08}, who find optically blue cores only in clusters where the central entropy is below 30keV cm$^2$, and \citet{Cavagnolo+08} who find that strong H$\alpha$ and radio emission is only present in BCGs below this same entropy. Enhanced H$\alpha$ and radio emission below this critical entropy suggests that both star formation and AGN activity in BCGs result from the same source, specifically cooled intracluster gas. This is further supported by the strong observational connection between optical lines in BCGs and the presence of colder gas phases traced by CO (\citealt{Edge+01,Salome+03}), atomic lines in the FIR (\citealt{Edge+10,Mittal+12}), warm molecular gas (\citealt{Edge+02,Egami+06,Donahue+11}) or dust (\citealt{Edge+99,Rawle+12}).
\\ \indent The location of the BCG relative to the X-ray emission also appears to be important in terms of BCG activity, where throughout the paper we use the term ``activity'' to refer to either, or both, star formation and AGN activity, unless otherwise noted. In \cite{Stott+12} for example a correlation between the radio loud fraction of BCGs and proximity to X-ray centroid is seen. \cite{Crawford+99} find that BCGs with emission lines have smaller separations between their position and the X-ray centroid than BCGs without lines. Similarly  \citet{Sanderson+09} find that all the LoCUSS BCGs with optical emission-lines are within 15 kpc of the X-ray centroid and a close correspondence between H$\alpha$ and radio emission and BCG/X-ray offset exists. These observations hint toward a strong association between activity in the BCG and its proximity to the centre of the cluster potential. 
\\ \indent The primary goal of this paper is to use optical, Mid-IR and UV survey data to investigate what extent correlations exist between BCG photometric colours and BCG activity. These correlations may be manifested as colour deviations from those of the bulk population of passive BCGs. 
With the advantage of a large sample of $\sim1000$ clusters, the aim is to investigate what proportion of BCGs, selected on host cluster X-ray emission alone, show signs of activity and determine how effectively we can detect active BCGs through photometry. In addition, we explore how, if at all, this relates to the overall X-ray properties of the cluster.
\\ \indent The organisation of this paper is as follows: we introduce the cluster sample and details of the photometric surveys in Section \ref{sect:data}. We present our analysis, results and discussion in Section \ref{sect:results} and conclude with a summary of the main results in Section \ref{sect:conclude}. Throughout this paper we use AB magnitudes, except for the \wise data which uses Vega, and assume a standard cosmology of $H_0 = 72\, \mathrm{km\,s^{-1}\,Mpc^{-1}}$, $\Omega_M = 0.27$ and $\Omega_{\Lambda}=0.73$.

\begin{figure}
\centering
\includegraphics[width=0.475\textwidth]{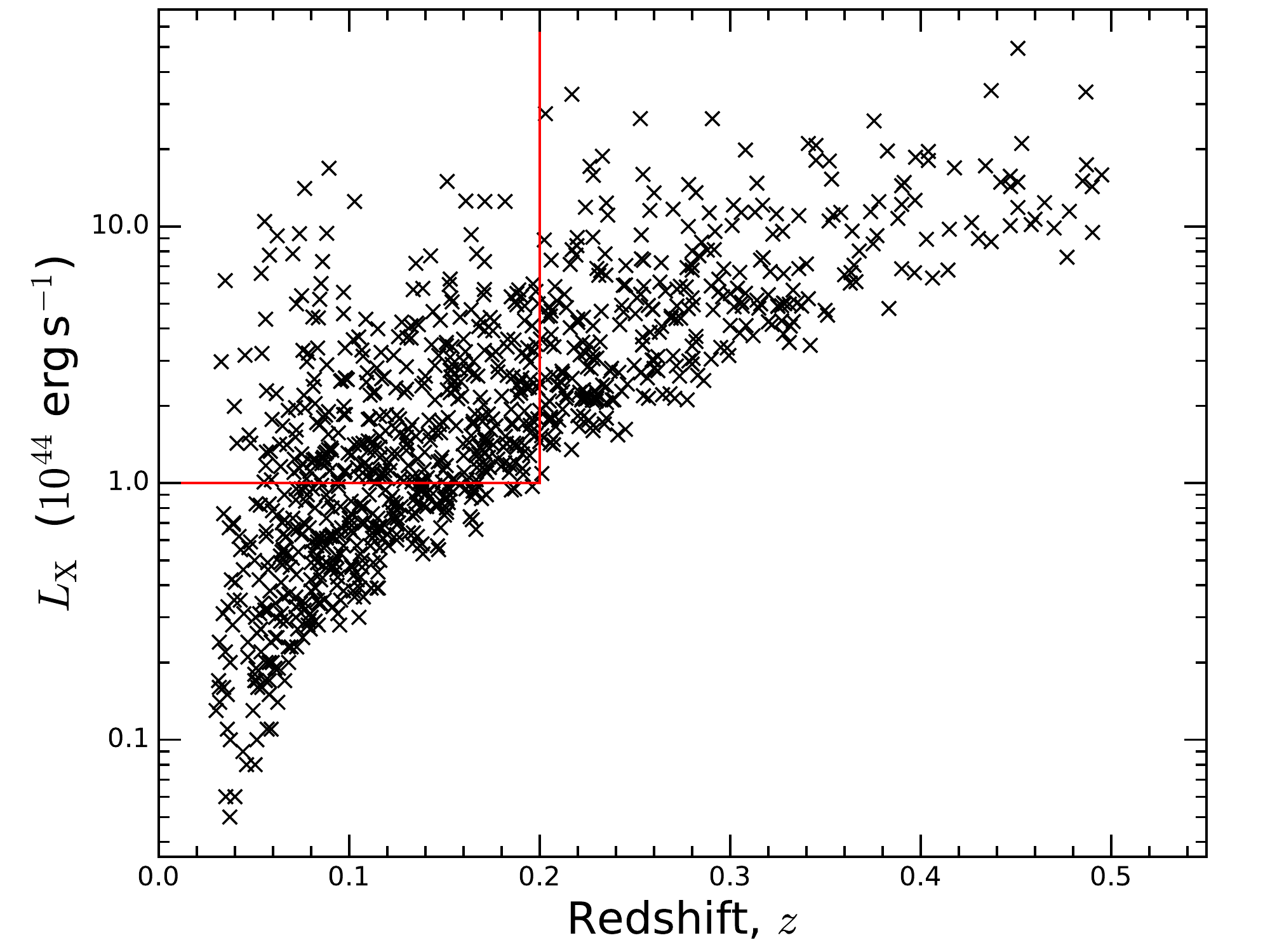}
\caption{The RASS 0.1--2.4~keV X-ray luminosity and redshift for each of the \TotalSample clusters in this sample. The solid (red) lines enclose the clusters selected in our `X-ray luminosity complete' subsample.}
\label{fig:Lx_z}
\end{figure}

\section{The Data}\label{sect:data}

\subsection{The Cluster Sample}
Our cluster sample consists of \TotalSample X-ray selected clusters between $0.03<z<0.5$. The sample was drawn from a systematic investigation into the \textit{ROSAT} All-Sky Survey (RASS) Bright Source Catalogue, BSC \citep{Voges+99}. Included in this are all published samples within the PanSTARRS-1 3$\pi$ footprint, namely the Brightest Cluster Sample, BCS, \citep{Ebeling+98}, the extended Brightest Cluster Sample, eBCS, \citep{Ebeling+00}, the ROSAT ESO Flux Limited X-ray survey, REFLEX \citep{Bohringer+04}, the Northern ROSAT All-Sky Galaxy Cluster Survey, NORAS, \citep{Bohringer+00} and the Massive Cluster Survey, MACS, (\citealt{MACS,Ebeling+07,Ebeling+10}). We have also included all cluster identifications of RASS Bright Source Catalogue sources below the flux limits of these published surveys. The sky coverage of these additional clusters is not uniform due to variation in the exposure time and Galactic column density, but this spatial incompleteness does not affect any particular class of cluster more than another. Therefore, while the fainter clusters are incomplete and do not constitute a flux-limited sample, we can still use these clusters as a fair sample of the overall population. The relatively short duty cycle of an AGN, or a burst of star-formation activity, within a BCG means that we need to maximise the parent sample we study in order to capture a meaningful number of these rare events. With X-ray luminosities ranging over two dex, but with a median \Lx$\sim2\times 10^{44}\,\mathrm{erg\,s}^{-1}$, the sample covers a broad range of cluster mass, whilst still including only massive clusters (i.e. not groups). The redshift and X-ray luminosity distribution of our sample is given in Fig. \ref{fig:Lx_z}. The X-ray luminosities are all drawn from the RASS, corrected for Galactic absorption and are quoted in the band 0.1--2.4~keV. We stress that from Fig. \ref{fig:Lx_z} it is evident that the highest X-ray luminosity clusters in our sample cover the full redshift range.

\subsection{X-ray Photometry: RASS}
The count rates were recorded by the PSPC in channels 11 to 235 (``broad band") 
during the RASS. These were then converted to unabsorbed X-ray fluxes in the 0.1--2.4 keV 
band by convolving the cluster X-ray emission with the PSPC detector
response function. In this process, the cluster emission is modelled as a
hot thermal plasma, characterised by metallicity (frozen at 0.3 Solar) and an
intra-cluster gas temperature k$T$. Since the latter has not been measured
for the vast majority of the clusters in our sample, it was estimated
iteratively, from the observed redshift and the inferred bolometric X-ray
luminosity in the cluster rest-frame, using the $L_{\rm X}$-k$T$ relation
of \cite{White+97}. In order to correct for absorption by Galactic gas we adopt the $n_{\rm H}$ column densities compiled by 
\cite{Dickey+Lockman90}.
In practice, this process for translating counts to fluxes is performed as an
interpolation in three dimensions ($n_{\rm H}$, k$T$, $z$).
\\
\indent We have included all cluster identifications of RASS Bright Source Catalogue \citep{Voges+99} sources below the flux limits of the published RASS cluster catalogues. However, as the RASS Bright Source Catalogue source detection was optimised for point sources, the count rate estimates from it may underestimate the counts from extended sources. The magnitude of this effect was investigated by \cite{Ebeling+98} as a function of extent and found to be significant (a factor of $>1.3$) in the most extended sources. We therefore make a statistical correction for the missing extended flux from the unpublished clusters by determining the ratio of count rates for the brightest clusters (the ratio of BCS and eBCS rates to the BSC rates) as a function of redshift.
Above our redshift lower limit of $z=0.03$ this ratio is at most a factor of 2.5, with
an average of 1.2 for $z>0.05$. The X-ray morphology of each individual cluster however is not accounted for in this correction, 
hence the flux of a few of the most nearby, most extended systems is likely underestimated.

\subsection{Spectroscopic Data}
The optical spectroscopy for the BCGs in the sample are drawn from a wide range of sources. The bulk of this comes from the nearly complete follow-up of the BCG sample by \cite{Crawford+99}, the large spectroscopic surveys by the Sloan Digital Sky Survey \citep{York+00} and the 6dF Galaxy Redshift Survey \citep{Jones+04}, a VLT FORS survey of BCGs in REFLEX (Edge et al., in prep) and any other spectra in the literature. In total, we have reliable spectra, sufficient to detect optical emission lines, for 73\% of the sample.

\subsection{Optical Photometry: Pan-STARRS}
The Pan-STARRS, PS1 $3\pi$ survey \citep{Tonry+12} is a wide-field photometric survey in the optical $g,r,i,z$ and $y$ bands. Covering the entire sky north of a Declination of $-30\degr$, the survey is conducted with the $1.8\,\mathrm{m}$ PS1 telescope in Hawaii, with an imaging pixel resolution of $0.25\arcsec$ per pixel. Using postage stamps from the PV3 data release, we performed aperture photometry on the \textit{griz} bands using SExtractor \citep{Bertin+96} - running in dual mode with the \textit{i}-band as the detection image (due to red-sensitive nature of the instrument). The inbuilt CLASS\_STAR function was used for star-galaxy separation with a very conservative cut of $<0.95$ applied, (where 1 is a star and 0 is not), as the misidentification of faint galaxies as stars was more problematic than stellar contamination in this work. Magnitudes are given by the MAG\_AUTO parameter, which measures the flux within a flexible elliptical aperture with a Kron radius \citep{Kron80}, and colours are derived from the MAG\_APER parameter, with a fixed circular aperture. 
\\ \indent An aperture diameter of 4 arcsec was used to measure colours. This is intentionally small, minimising contamination from source blending in the dense cluster environment. This is supported by a $\chi^2$ test which revealed apertures of this size minimised the scatter in the colour-magnitude relation. Metallicity (or age) induced colour gradients across galaxies are known to exist \citep{Cardiel+98}, so in order to quantify the effect of aperture size on colours, the 40 lowest redshift BCGs (median $z = 0.035$), which generally have the largest angular size on the sky and hence should represent the most extreme cases, were measured at aperture diameters of 4, 15, and 30 arcsec. In the case of the most extreme difference, that is between the (\grb) colour measured in a $30\arcsec$ and $4\arcsec$ apertures, (corresponding to median angular diameters of $20.5\,\mathrm{kpc}$ and $2.7\mathrm{kpc}$ respectively), a median/mean colour difference of only $0.04$ mags was measured, which has a negligible effect on our analysis.
\\ \indent Throughout this paper Galactic reddening corrections are made using the Galactic Extinction calculator available through the NED\footnote{http://ned.ipac.caltech.edu/forms/calculator.html}, based on the \cite{Schlafly+11} extinction maps. And where needed, K-corrections are made assuming a simple stellar population model, \citep{Bruzual+03}, with solar metallicity, a Chabrier Intial Mass Function, formation at $z=3$ and subsequent passive evolution.

\subsubsection{Identifying the BCG}\label{S:BCGident}
The identification of the BCGs was made via a visual inspection in the Pan-STARRS $3\pi$ imaging. The selection was primarily based on their morphology and centrality, looking for the visually most extended galaxy - often with a cD-like profile - at the centre of the galactic distribution. As such, the BCG need not necessarily be photometrically the brightest galaxy. When the choice of galaxy was ambiguous, the candidate best aligned with the X-ray emission was selected. Visual identification, although time consuming, is probably still the most reliable method, minimising issues with projection. Visually inspecting each image also aided in the identification of misclassified X-ray point sources in the sample, since these generally lack the visually obvious galactic overdensities characteristic of clusters.
\\ \indent \citet{Lauer+14} find that if taking the literal definition of a BCG as the brightest galaxy, then $12\%$ of local clusters have a BCG $>500\,\mathrm{kpc}$ from the X-ray centre. They conclude these are likely drawn into a cluster from recent merger events, as modelled by \citet{Martel+14}. However, as we are interested in the interplay between the BCG and its traditionally-central environment, we redefine our BCGs as the brightest, most extended, central galaxy.

\begin{figure}
\centering
\includegraphics[width=\columnwidth]{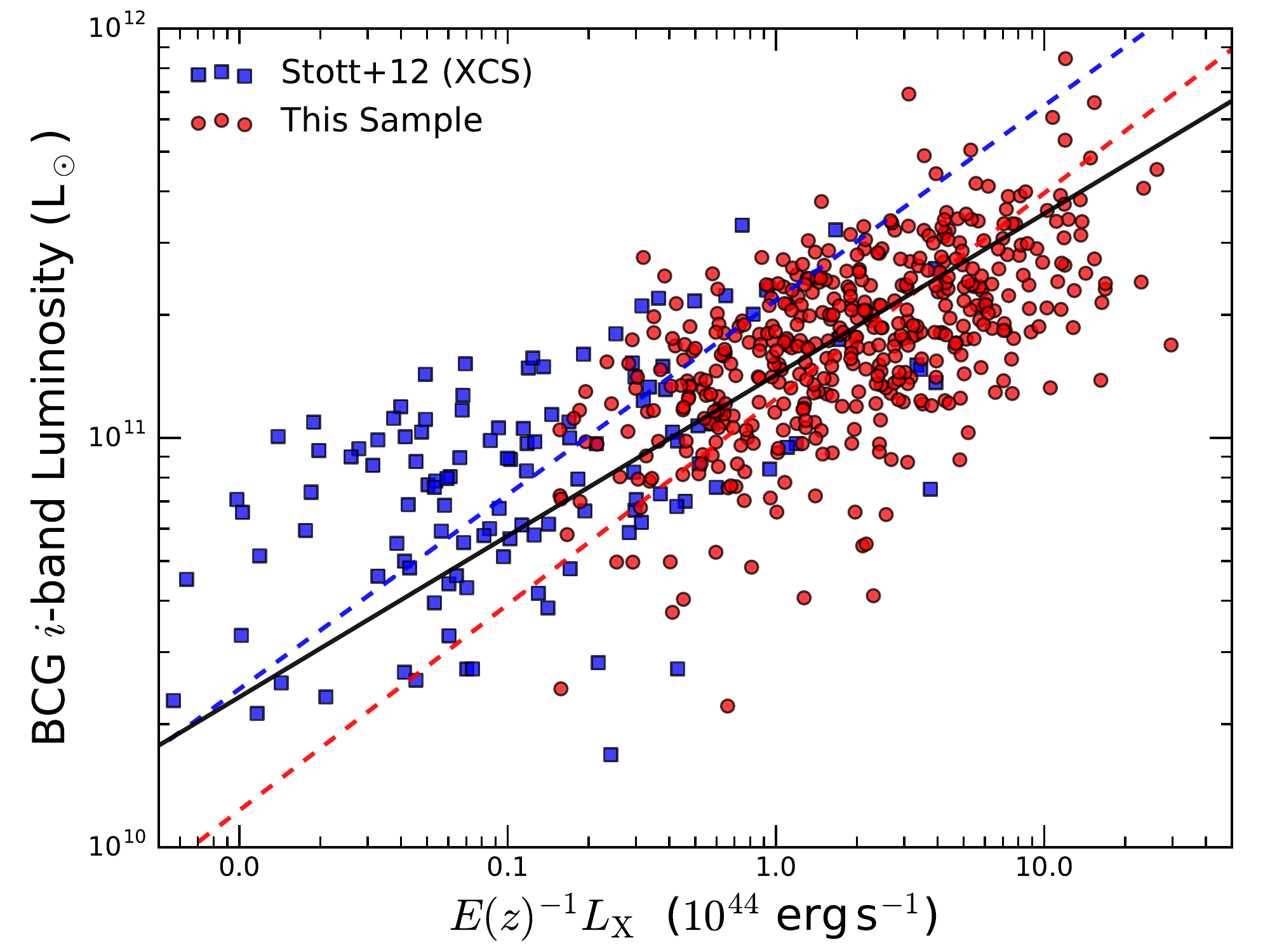}
\caption{The \textit{i}-band luminosity of BCGs, against the 0.1--2.4 keV X-ray luminosity of the host cluster. 
The (red) circles represent our cluster sample and the (blue) squares represent the groups/clusters from Stott et al. (2012) - corrected to have optical and X-ray measurements comparable to our own. The dashed lines represent the BCES bisector best fits to the sample of corresponding colour and the solid black line represents a BCES bisector fit to the combined data sets.}
\label{fig:Lx_BCGi_s12}
\end{figure}

\begin{figure}
 \resizebox{\columnwidth}{!}
 {\includegraphics{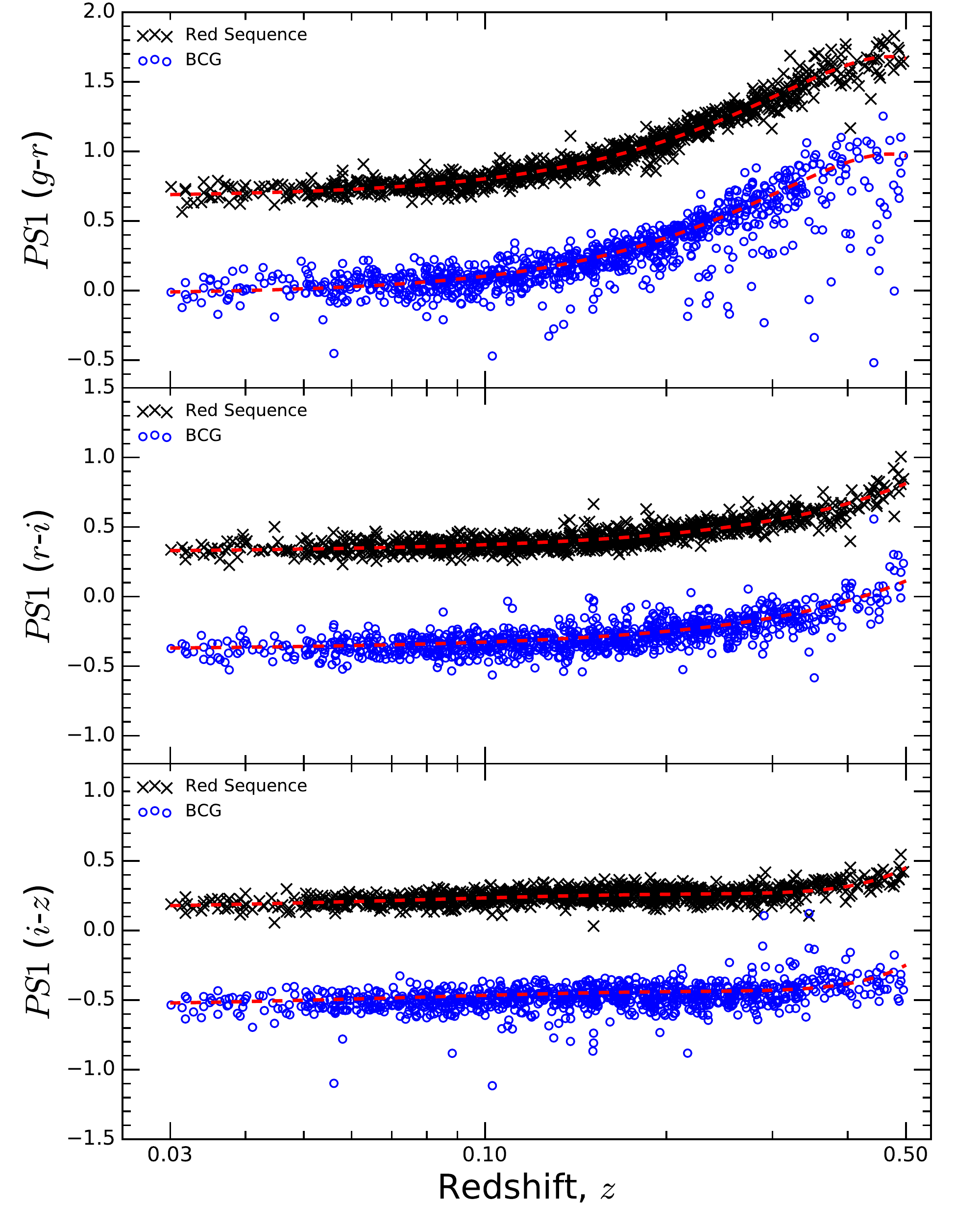}}
 \caption{PS1 \grb (top panel), \rib (middle panel) and \izb (bottom panel) colours, against redshift. The (black) crosses show the colour of the cluster red sequence, measured at a flux of 19th magnitude for each cluster. The (blue) empty circles show the colour of the BCG for each cluster, but corrected to be at a flux of 19th magnitude also. The BCGs (empty circles) have been artificially offset by $-0.7$ mags for visual clarity. The upper (red) dashed line in each panel shows a best fit to the red sequence. These exact same lines are offset by $-0.7$ mag also, forming the lower dashed lines, this is to show the reader that the bulk population of BCGs lie on the cluster red sequence as expected for a passive galaxy. 
}
 \label{Fig:logRSBCGz}
\end{figure}
 
\subsection{Mid-IR Photometry: \textit{WISE}}
The Wide-field Infrared Survey Explorer (\textit{WISE}) \citep{WISE10} has covered the whole sky in the  Mid-IR at $3.4$, $4.6$, $12$ and $22\,\micron$. From the AllWISE Source Catalogue we extract photometry for our BCG sample in the \wb1, \wb2 and \wb3 bands at $3.4$, $4.6$ and $12\,\micron$ respectively. The \wise source closest to the BCG position is selected, which is within $1\arcsec$ for $79\%$ of BCGs, $2\arcsec$ for $94\%$ of BCGs and within $6\arcsec$ for all BCGs. We do not perform any Galactic extinction corrections as this is negligible for Mid-IR observations \citep{Cardelli+89}. 
\\ \indent The full BCG sample is well detected in \wb1 and \wb2, with a Signal to Noise, $S/N>\SNfit$ for all observations. At $12\,\micron$ (\wb3) however some sources are not detected due to the flux limit of \wise. We thus restrict our \wb3 analysis to detections with $S/N>\SNfit$ and find that the number of robustly detected BCGs in \wb3 is \WISESNnum ($56\%$ of total sample size).  

\subsection{UV Photometry: \textit{GALEX}}
The Galaxy Evolution Explorer (\galex) \citep{GALEX05} is a space based observatory imaging in two ultraviolet bands, the Far-UV (FUV) at $1350$--$1780$\r{A} and the Near-UV (NUV) at $1770$--$2730$\r{A}. We used the aperture photometry available from the \textit{GALEX} catalogue\footnote{http://galex.stsci.edu/GalexView/} and extracted any source within six arcsecs of our BCG position. If multiple observations were returned, we attempted to maximise the signal to noise by taking the observation with the longest exposure time - unless the next longest exposure was more central and within 2/3 of the highest exposure time, since the PSF is poorer with radius from the centre in the \galex field of view. We correct for Galactic extinction assuming $A_{\mathrm{FUV}}=2.5\,A_V$ and $A_{\mathrm{NUV}}=3.25\,A_V$ \citep{Hoffer+12}.
\\ \indent The \galex catalogue is compiled of observations from numerous nested \galex surveys, which vary in sky coverage and depth, none of which are full sky and hence our sample is far from complete in the UV. The widest survey, the All-sky Imaging (AIS), (which despite the name covers only 26,000 square degrees of the sky) has typical exposure times of only 100s, which are insufficient for a robust detection for most of our redshift range. Hence, despite the catalogue returning a detection for 541 of our BCGs within 6\arcsec, when we restrict our analysis to detections with $S/N>\SNfit$, the number of galaxies reduces to \NUVrob BCGs ($\sim25\%$ of the total sample).  

\subsection{Radio: NVSS}
The NRAO VLA Sky Survey (NVSS) \citep{NVSS98}, operating at $1.4,\mathrm{GHz}$, covered the entire sky north of a Declination of $-40\degr$ and thus provides observations of our entire sample. We use the NVSS Source catalogue\footnote{www.mrao.cam.ac.uk/projects/surveys/nrao/NVSS/NVSS.html} to extract the photometry for the nearest detection within $45\arcsec$ of the BCG position. Whilst the spatial resolution of NVSS does not permit us to attribute a radio detection to the BCG with certainty, BCGs are the most likely cluster galaxy to host a radio-loud AGN (\citealt{Burns+81,Valentijn+83,Burns+90,Best+07}) and hence we assume any radio detection to be associated with the BCG. We find \RadioDet BCGs with an associated radio detection, corresponding to a radio detection rate of $52\%$, this is comparable to the $\sim60\%$ detection rate of \cite{Hogan+15}, but, logically, less given our sample goes to higher redshift.

\begin{figure}
\begin{minipage}{\textwidth}
 {\includegraphics[width = 0.475\textwidth]{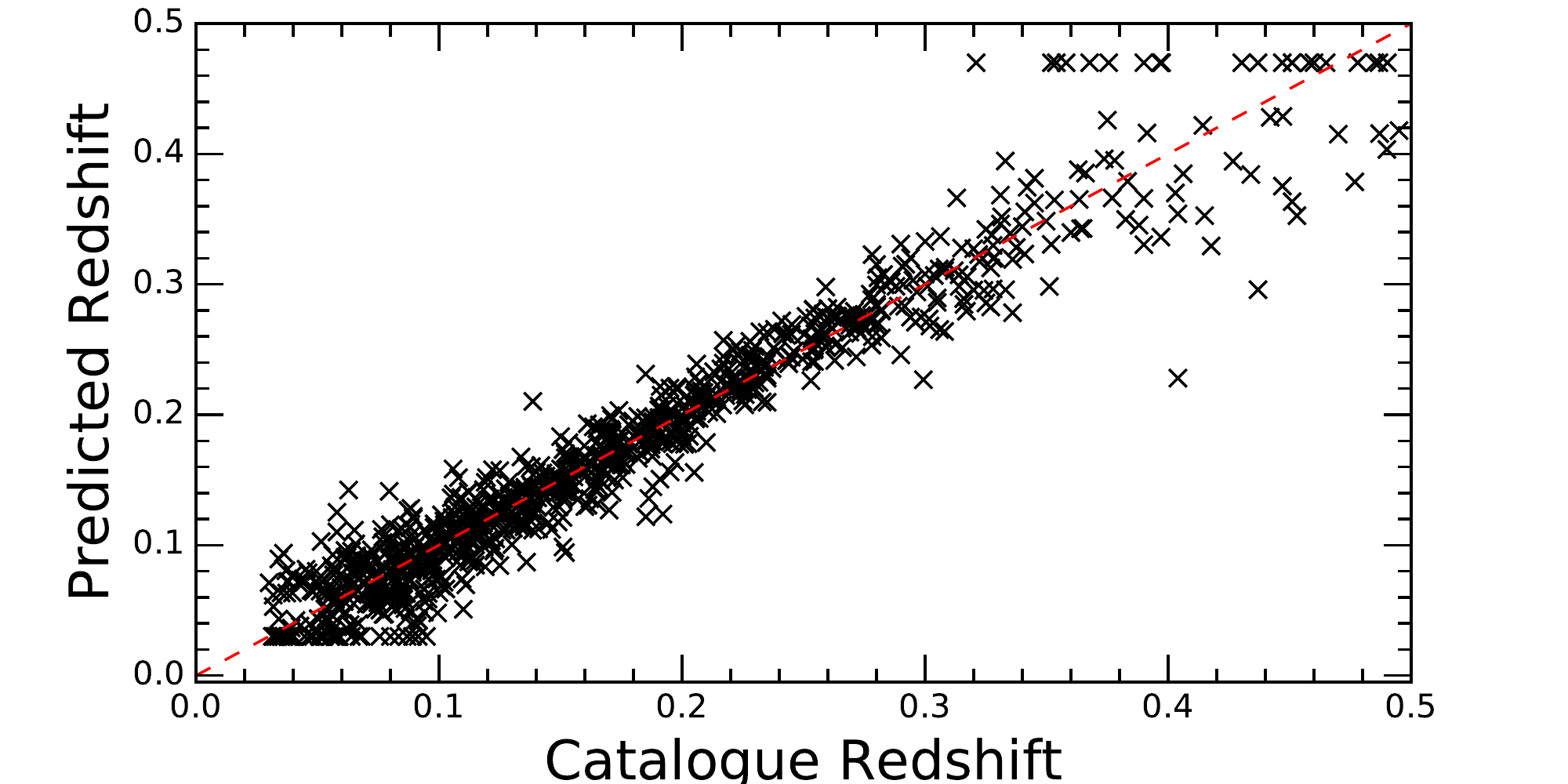}}

\vspace{0.3cm}
 {\includegraphics[width = 0.475\textwidth]{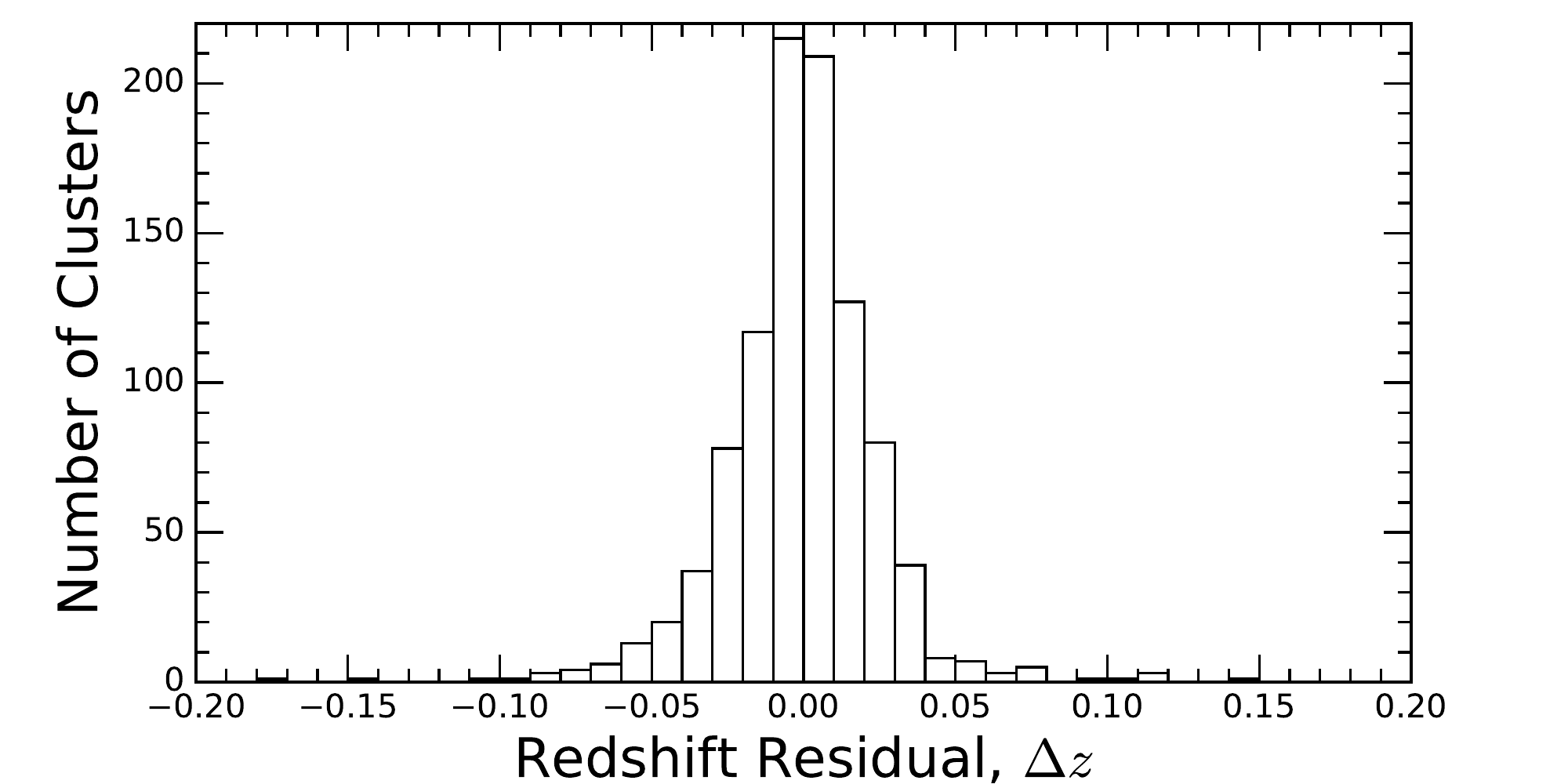}}
 \end{minipage}
\caption{Top: The predicted redshift, derived from the  \grb colour of the red sequence, against the catalogued redshift for each cluster. Bottom: The redshift residuals between the predicted and actual redshifts. The distribution has a mean at zero and standard deviation of $0.025\,$. At higher redshifts there are sometimes very few red sequence galaxies above the PS1 flux depth, and as a result the true red sequence is difficult to select correctly; this explains those cases where the predicted redshift and actual redshift significantly deviate. Also, any red sequence colour above our peak in the curve results in a predicted redshift corresponding to the sample maximum at $\sim0.46$.}
\label{Fig:zAnal}
\end{figure}


\section{Results}\label{sect:results}

\subsection{Optical Analysis and Results}

\subsubsection{BCG Luminosity}
We begin our optical analysis by comparing the properties of the BCG with those of the host cluster, as seen in Fig. \ref{fig:Lx_BCGi_s12}. We find that the BCG \ib -band luminosity is weakly correlated with the host cluster X-ray luminosity, such that, whilst for any individual BCG the scatter is significant, we do see a general trend in which the brighter BCGs tend to reside in the more X-ray luminous clusters. This reinforces previous results showing a correlation between optical BCG luminosity and cluster X-ray luminosity (e.g. \citealt{Edge91, Stott+12}). We conclude that, given that optical luminosity traces stellar mass and cluster X-ray luminosity traces the cluster mass, the most massive BCGs tend to belong to more massive clusters. This interpretation is supported by the results of \cite{Stott+10} and \cite{Lidman+12} who derive BCG stellar masses and cluster masses from BCG Near-IR luminosities and cluster X-ray temperatures respectively, (which are both more reliable tracers of mass). Additionally, a positive correlation between BCG luminosity and the cluster velocity dispersion -- another tracer of cluster mass -- has also been observed (\citealt{Whiley+08,Lauer+14}).

\indent In Fig. \ref{fig:Lx_BCGi_s12} we plot the corrected values of \Lx and $L_{\mathrm{i}}$ from \cite{Stott+12} (hereafter S12) alongside our own for comparison. The S12 $L_{\mathrm{i}}$ values are derived from SDSS model magnitudes, which have systematically brighter fluxes than measured through our own PS1 photometry by a few tenths of a magnitude. Since this offset in BCG flux is uniform with redshift we apply a correction of the median value of 0.4 mags to the S12 values. Additionally we apply a correction in the S12 \Lx values, (taken from the XCS \citep{Mehrtens+12}), from bolometric values to that of 0.1--2.4 keV. This is achieved using the ratio of 0.1--2.4 keV flux to bolometric flux, as a function of temperature, and the XCS temperature values from S12. Applying a linear regression BCES bisector fit \citep{Isobe+90} to the points we find, \begin{equation} \mathrm{log}_{10}(L_i) = \alpha \mathrm{log}_{10}(E(z)^{-1}L_{\mathrm{X}})+\beta, \end{equation} where $L_i$ is the BCG \ib -band luminosity (L\sun), \Lx is the 0.1--2.4 keV X-ray luminosity ($10^{44}\,\mathrm{erg\,s}^{-1}$) and $E(z) = [\Omega_m(1+z)^3+\Omega_{\Lambda}]^{1/2}$ with $\alpha = 0.48\pm0.08$ and $\beta = 11.34\pm0.08$ for the S12 sample, $\alpha = 0.50\pm0.04$ and $\beta = 11.09\pm0.02$ our sample and $\alpha = 0.39\pm0.04$ and $\beta = 11.15\pm0.09$ for the combined samples. Hence we find, when comparing the best fitting lines to this sample and the S12 sample, that the slopes are consistent but there is an offset in the normalisation. The interpretation is that this reflects the differing sample selection -- specifically that the XCS sample is dominated by low mass galaxy groups/clusters and our sample is dominated by massive clusters. We find the slope for the combined samples is flatter than either separate sample, suggesting a possible flattening of the relation in low mass halos. This is perhaps driven by a lower prevalence of AGN feedback in low mass halos. Hence the stellar mass growth of their central galaxies is not as effectively truncated. 

\subsubsection{Cluster Red Sequence}
\indent The cluster red sequence, the observed linear relationship between magnitude and colour for cluster galaxies, was selected using a multi-colour band approach. Firstly, a conservative colour cut is applied around the expected red sequence colour given its redshift. Then magnitudes are binned and the median colour selected for each bin. A further colour cut is applied around these median colours and a linear fit is then made to these sources, constituting the initial red sequence fit. The colour index spanning the rest-frame $4000$\r{A} break and the adjacent redder-colour index are then cross-matched, where only sources that fit the initial red sequence in both colour bands are classified as red sequence galaxies. Because cluster early type galaxies are the reddest, brightest galaxies at a given redshift, this stage reduces contamination of the red sequence as spanning the 4000\r{A} break provides the most significant contrast with field galaxies. The linear fit and cross-matching process is then repeated iteratively, decreasing the colour clipping each time, until a final clip of $\pm0.1\,\mathrm{mags}$ around the red sequence best fit is made and defined as our cluster red sequence. Our analysis of the colour-magnitude relation is restricted to a radius of $0.5\,\Mpc$ around the BCG position. 
\\ \indent In Fig. \ref{Fig:logRSBCGz} we explore the PS1 \grb, \rib and \izb colours of the red sequence, measured at a fixed flux of 19th magnitude and how this relates to redshift. We measure this at magnitude 19 as the red sequence generally pivots about this value, thus minimising the colour error due to the uncertainty of the red sequence slope. The small level of scatter of this distribution illustrates the homogeneity of the cluster red sequence population, despite the different cluster environments and masses in our sample \citep{Andreon03}. The observed evolution with $z$ is dominated by the transition of the $4000$\r{A} break through the photometric bands (i.e. the redshift evolution that one would K-correct for), and also the relative change in position of 19th magnitude along the red sequence gradient (which typically goes redder at the bright end). 
\\ \indent The tight relationship in red sequence colour with redshift means the colour of a cluster's red sequence can provide an estimate for its redshift. In Fig. \ref{Fig:zAnal} we test this by comparing the redshift predicted by its red sequence \grb colour, against the actual redshift, for our cluster sample. We find that the predicted and actual redshift agree to within $0.025$ at $1\,\sigma$. The increased scatter at higher $z$ is due to the increased difficulty in reliably selecting the red sequence. At high redshift, ($z>0.3$), only the very brightest few red sequence galaxies are detected, giving few points on which to select and extrapolate the red sequence. The largest source of uncertainty is correctly sampling the red sequence slope, which is used when evaluating the \grb offset. Poor selection of the red sequence at high $z$ accounts for those clusters with significant redshift residuals. 
\\ \indent The colour of the cluster red sequence provides a standard for the colour one would expect for a passively evolving galaxy on the red sequence. The equivalent colour of the BCG can thus be compared to this and hence used to determine whether deviations from passivity exist for these galaxies. In addition to this, analysis of red sequence was a useful tool in identifying and removing any misclassified source in the X-ray catalogue. For example, a number of sources in the initial sample were point sources, such as AGN and white dwarfs. The lack of red sequence in such cases made identification, and thus removal, of these simple.   
\begin{figure}
\centering
 \resizebox{\columnwidth}{!}
 {\includegraphics{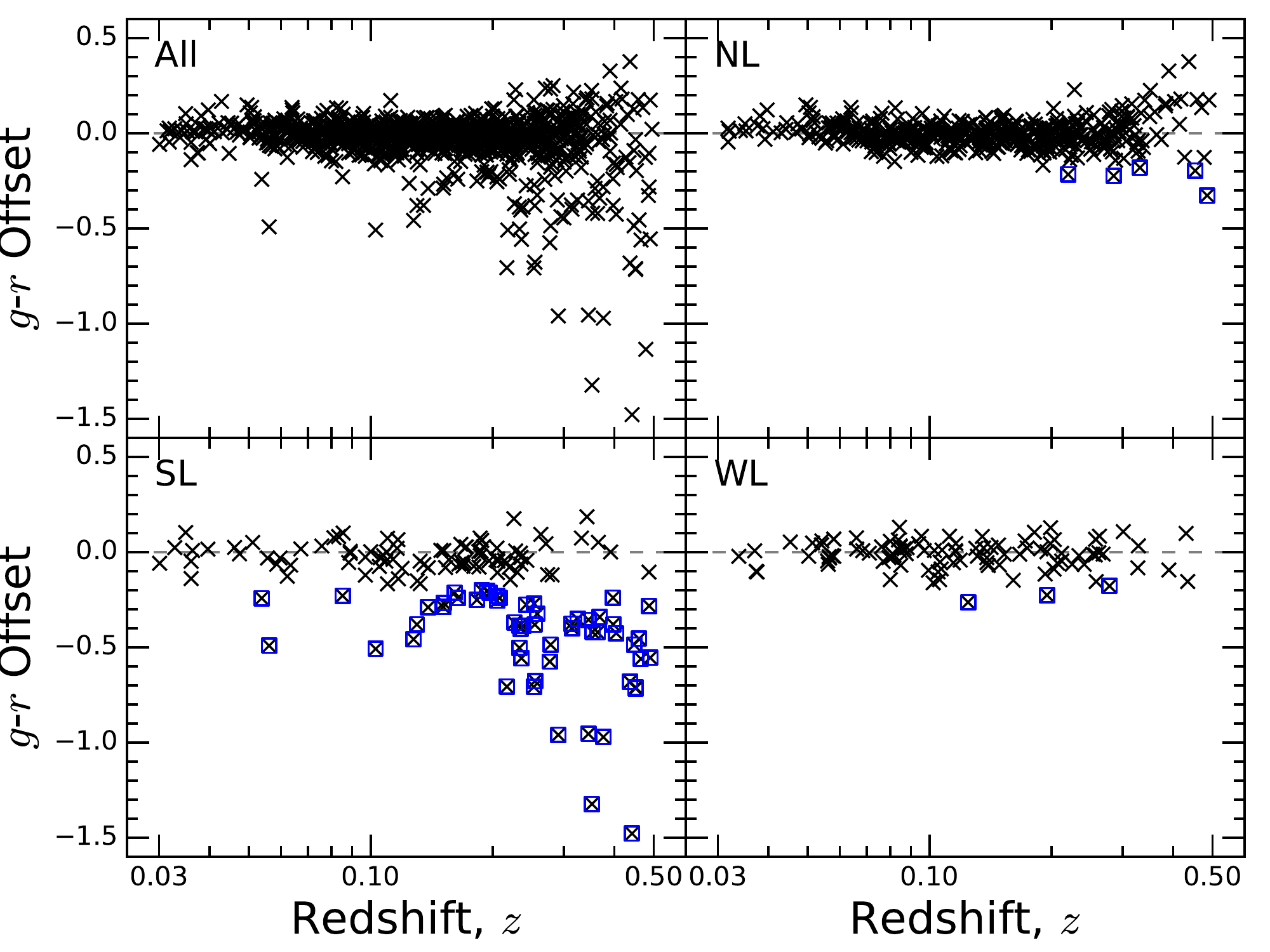}}
 \caption{Each panel shows the BCG \grb colour offset -- defined as the difference between the BCG and red sequence colour -- against redshift. \SubCapt}
\label{fig:grOffz}
\end{figure}

\begin{figure}
\centering
 \resizebox{\columnwidth}{!}
 {\includegraphics{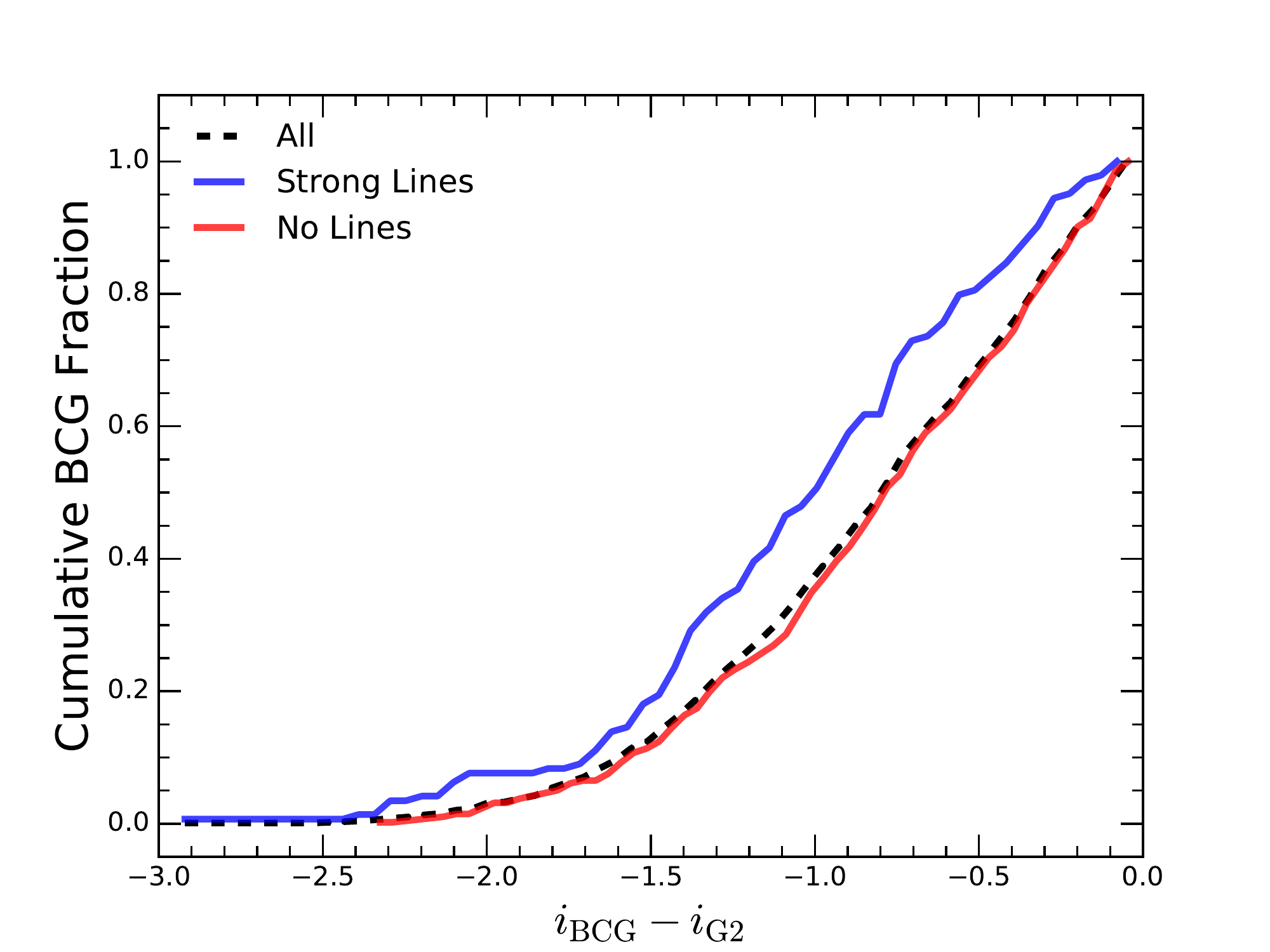}}
 \caption{The cumulative fraction of clusters as a function of the $i$--band magnitude difference between the BCG and \textit{G2}, the 2nd ranked cluster red sequence galaxy. 
}
\label{fig:magdif}
\end{figure}
\begin{figure}
\centering
\includegraphics[width=\columnwidth]{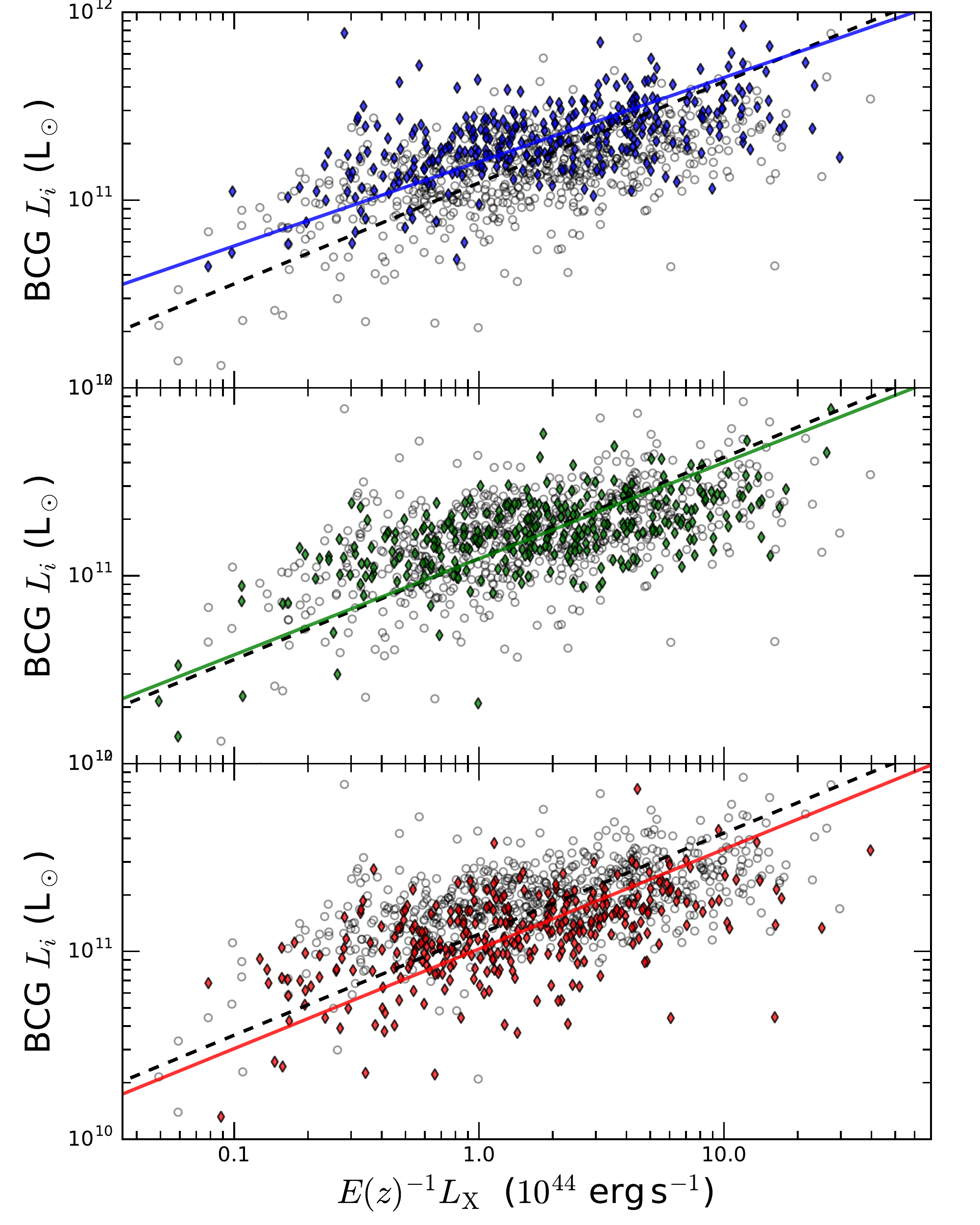}
\caption{The PS1 \textit{i}--band luminosity of the BCG, against the 0.1--2.4 keV X-ray luminosity of the host cluster.
The sample is split equally into three based on BCG dominance, described by the difference in magnitude between the BCG and the 2nd ranked galaxy. The filled diamonds in the top panel correspond to BCGs in the top tier -- that is, the BCGs with the biggest magnitude difference -- the filled diamonds in the second panel correspond to the middle tier and the filled diamonds in the bottom panel correspond to the bottom tier. The empty circles show the full sample to aid visual comparison. 
The dashed line indicates the best fit, made using a BCES bisector fit, to the full sample, the solid line in each panel indicates a best fit to the filled diamonds of that panel.
}
\label{fig:Lx_BCGi_m12}
\end{figure}

\subsubsection{BCG Colour}\label{Sect:Dom}
\indent In Fig. \ref{Fig:logRSBCGz} we also display the PS1 \grb, \rib and \izb colours of the BCG in each cluster, corrected to be at the same flux as the red sequence. We see that the BCG colour evolution generally follows the same trend with redshift as the red sequence, indicating that most BCGs lie on the red sequence. As with the red sequence colours, the BCGs form a fairly tight relation suggesting BCGs are generally a fairly homogeneous galaxy population. This is to be expected if you consider BCGs to be quiescent galaxies that formed their stellar populations at $z>2$, like the other cluster members. However, we see also a significant number of BCGs with optical colours that differ from that of the bulk, passive population, particularly in \grb.
\\ \indent In Fig. \ref{fig:grOffz} we show the \grb colour offset, defined as the colour difference between the BCG and the red sequence at the same flux. There are a number of BCGs which have \grb colours significantly bluer than the red sequence galaxies, which could be indicative of star formation and/or AGN activity. To explore this further we split our BCG sample by emission line status as both star formation and AGN activity lead to optical emission lines in galaxies. We divide our BCGs into: those with strong emission lines (H$\alpha$ slit flux $>10^{-15}$~erg~cm$^{-2}$s$^{-1}$), those with weak emission lines ($<10^{-15}$~erg~cm$^{-2}$s$^{-1}$), those with no lines, and those without any spectral data available. 
\\ \indent We see from Fig. \ref{fig:grOffz} there is a clear connection between \grb colour offset and line status with most blue BCGs also showing strong emission lines. This is in stark contrast to BCGs without emission lines, which have a mean \grb offset of zero, as expected for passive galaxies. If we thus assume the non line-emitting BCGs are representative of the quiescent population, we can use these to estimate the general scatter in colour and hence quantify what constitutes a significant deviation in colour from passivity. We find a standard deviation of $\sigma_p = 0.07$ mags for the non line-emitting (``passive'') BCGs. 
\\ \indent Defining our threshold for BCG activity as a colour offset in excess of $2.5\,\sigma_p$, we find that $8\%$ of BCGs exhibit significantly blue colours and as such are classed as ``active''. Of the strong line emitting subsample, $39\%$ of BCGs show a \grb offset $>2.5\,\sigma_p$ and the mean offset is $-0.19$ mags. In fact, all the BCGs with \grb offsets $<-0.5$ mags are either strong line emitters or have no available spectra (and are likely strong line emitters themselves). The strong connection between optical colour and line status is indicative that these BCGs are optically blue because of activity in the BCG. Note that we have manually investigated the colour magnitude diagrams of all apparently blue BCGs to rule out incorrect selection of the red sequence as an origin for the offset in \grb, in addition to double checking our BCG selection was correct. A summary of the fraction of optically blue BCGs as related to line status is provided in Table \ref{tab:fract}. 
\\ \indent It is interesting to note that whilst optical emission lines are indicative of activity, $\sim60\%$ of those BCGs with emission lines do not show up as optically blue. In most cases this is likely a result of the fact that obscured SF and/or AGN activity can lead to ionisation of the gas, visible as strong emission lines, whilst having minimal influence on the continuum emission, and hence colour, of the overall galaxy. Additionally, the optical line emission associated with the BCG may result, not from ionising stars or AGN, but from ionisation at the hot/cold interface between clouds of cold molecular gas and the surrounding hot ICM in the cores of cool core clusters (\citealt{Ferland+09,Fabian+11, Hamer+16}). Aperture effects may also be a contributing factor, whereby the photometry and spectroscopy are probing different regions of the BCG. As an aside, it should be noted that the fact some BCGs are optically blue has implications regarding optical cluster algorithms, which often select the BCG assuming a BCG colour on the red sequence.  

\subsubsection{BCG Dominance}
Next we investigate the connection between the emission line status of the BCG, and its dominance relative to its companions in the cluster. In Fig. \ref{fig:magdif} we show the cumulative fraction of BCGs, as a function of the magnitude difference between the BCG and the 2nd ranked galaxy on the red sequence, $G2$. A connection appears to exist between the luminosity dominance of the BCG and its emission line status, with a higher fraction of strong line emitting BCGs showing a large magnitude separation between itself and $G2$ compared to those with no lines. This observation is consistent with implications from the literature, that is, that BCGs with emission lines trace cool core clusters (\citealt{Heckman+81, Hu+85, Johnstone+87, Heckman+89, Crawford+99,  McDonald+11}), thus tend to be located at the very centre of the cluster potential \citep{Sanderson+09} and the observation that BCGs are more likely to be cD galaxies when in close proximity to the cluster centre \citep{Lauer+14}. This also agrees with \cite{Smith+10} which demonstrated a direct association between the BCG dominance and the central gas density profile of the cluster, specifically that clusters hosting a high luminosity gap also host strong cool cores.

\indent BCG dominance is also explored in Fig. \ref{fig:Lx_BCGi_m12} - here we divide the full sample into three based on the magnitude difference between themselves and $G2$ and find that the most dominant BCGs are systematically more luminous than other BCGs in clusters of similar X-ray luminosity, with the converse being true for the least dominant BCGs. This supports the notion that the magnitude difference is dominated by the brightness of BCGs themselves and not just reflective of a deficit in the brightness of $G2$ in these particular systems. In principle, the apparent correlation between BCG $L_{\mathrm{i}}$ and cluster \Lx that is seen could be explained by the association between cool core clusters and an enhanced BCG dominance - whereby the cool core nature of a cluster leads to an enhancement in both the $i$--band and X-ray brightness. However the existence of this correlation in all three subsamples, even in the lowest third by luminosity gap, supports the interpretation that this is not a significant factor here. 

\begin{figure}
\centering
 \resizebox{\columnwidth}{!}
 {\includegraphics{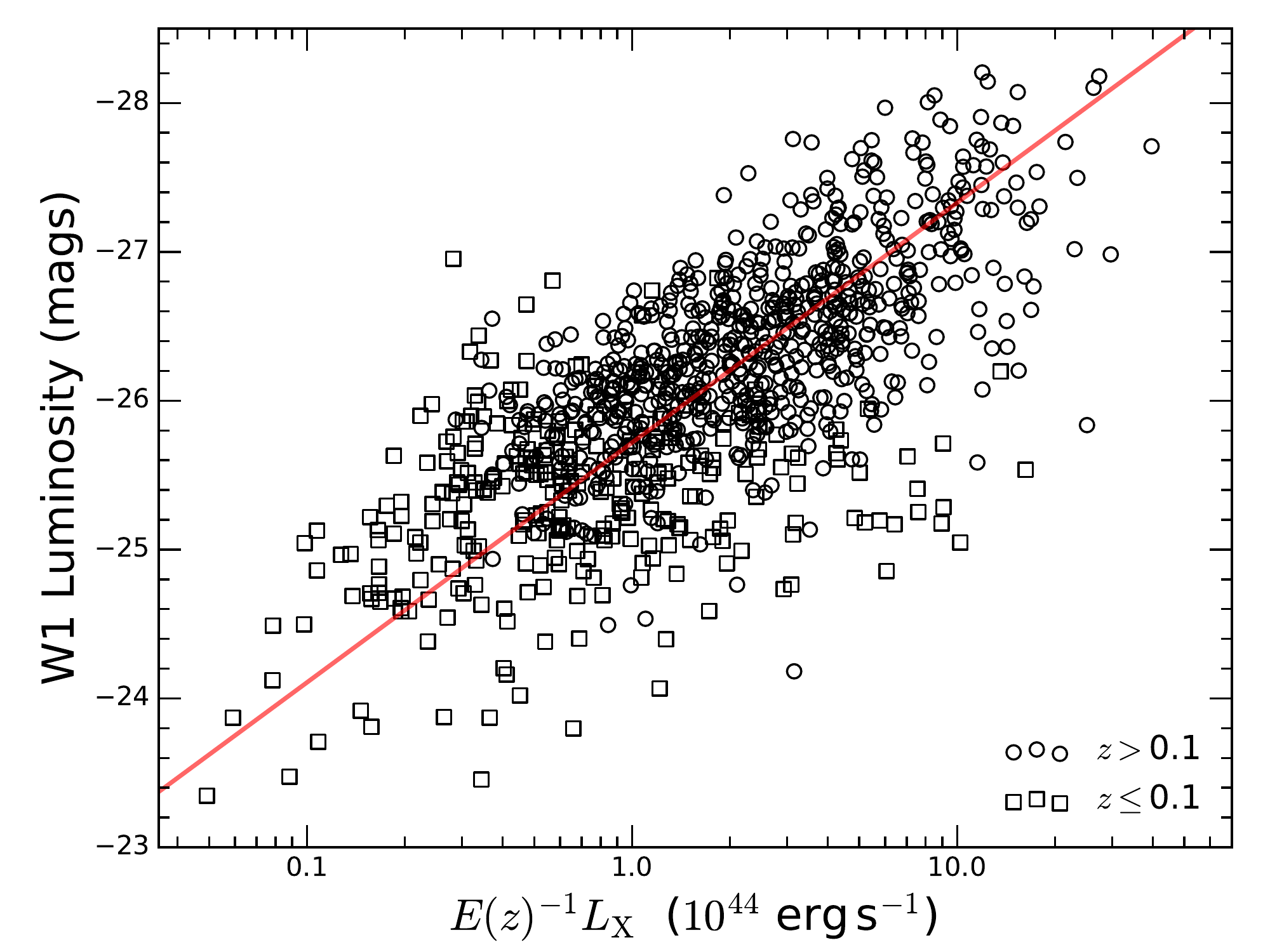}}
 \caption{The \wise \wb1 (3.4\micron) luminosity of the BCG against the 0.1--2.4 keV X-ray luminosity of the host cluster. 
The (red) dashed line indicates our best fit, made using a BCES bisector fit.
}.
 \label{Fig:w1Lx}
\end{figure}

\begin{figure}
\centering
 \resizebox{\columnwidth}{!}
 {\includegraphics{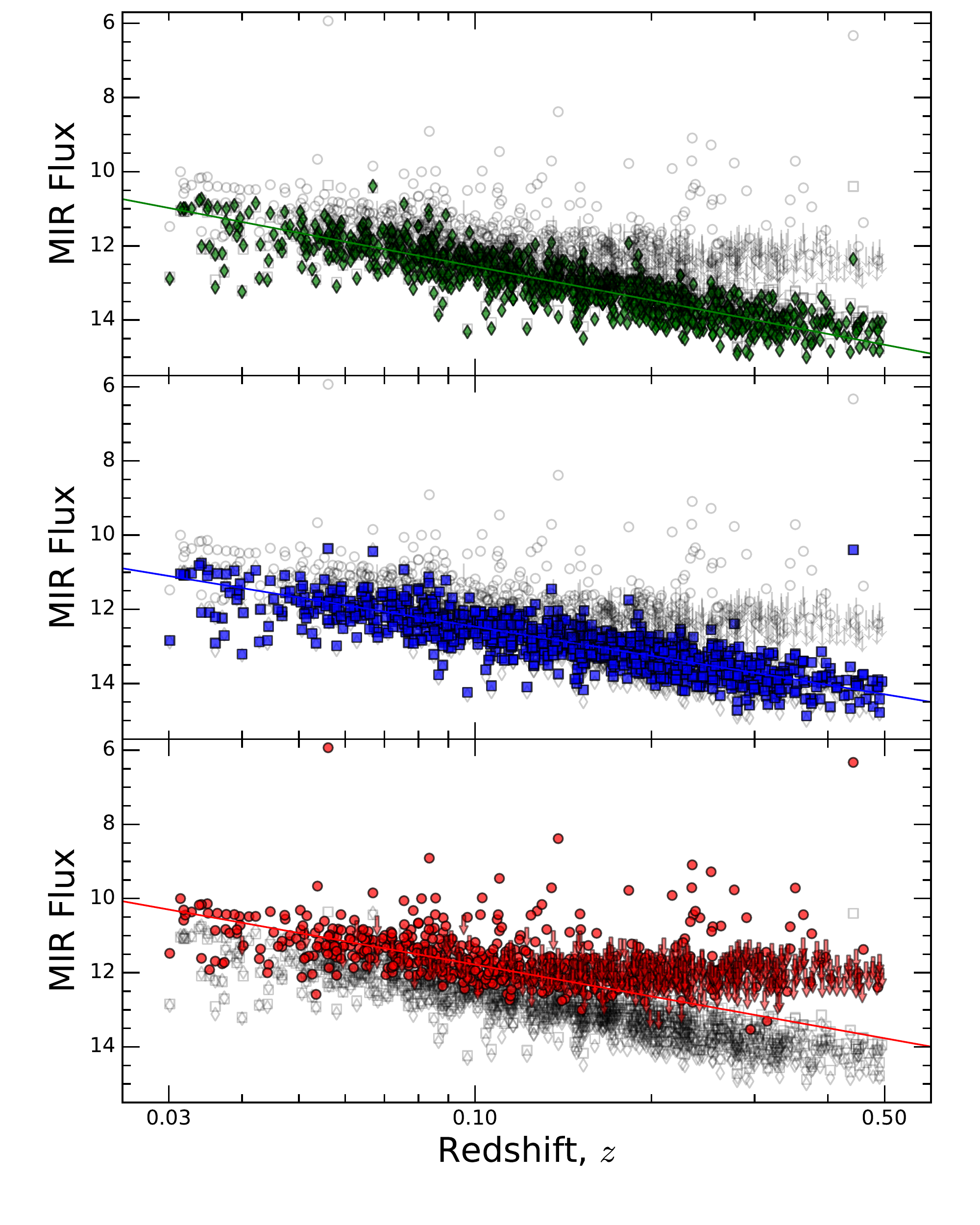}}
 \caption{The \wise fluxes of our BCGs against redshift. The \wb1 ($3.4\micron$) fluxes are given by the (green) diamonds in the top panel, the \wb2 ($4.6\micron$) fluxes by the (blue) squares in the middle panel and the \wb3 ($12\micron$) fluxes by the (red) circles in the bottom panel. The grey points correspond to the values of the other bands with their respective symbols. Any points given as arrows indicate limiting magnitudes with poor signal to noise ($S/N<$\SNfit). Clearly from this figure the depth of \wb3 is insufficient to probe the quiescent galaxy population beyond $z\sim0.1$, but crucially for our study, those with a strong \wb3 excess continue to show up at least until $z=0.5$. The solid lines show the respective best fit trends to the bulk quiescent population of BCGs, in the case of \wb3, this is extrapolated from $z\leq0.1$. }
 \label{Fig:flux-z}
\end{figure}

\begin{figure}
\centering
 \resizebox{\columnwidth}{!}
 {\includegraphics{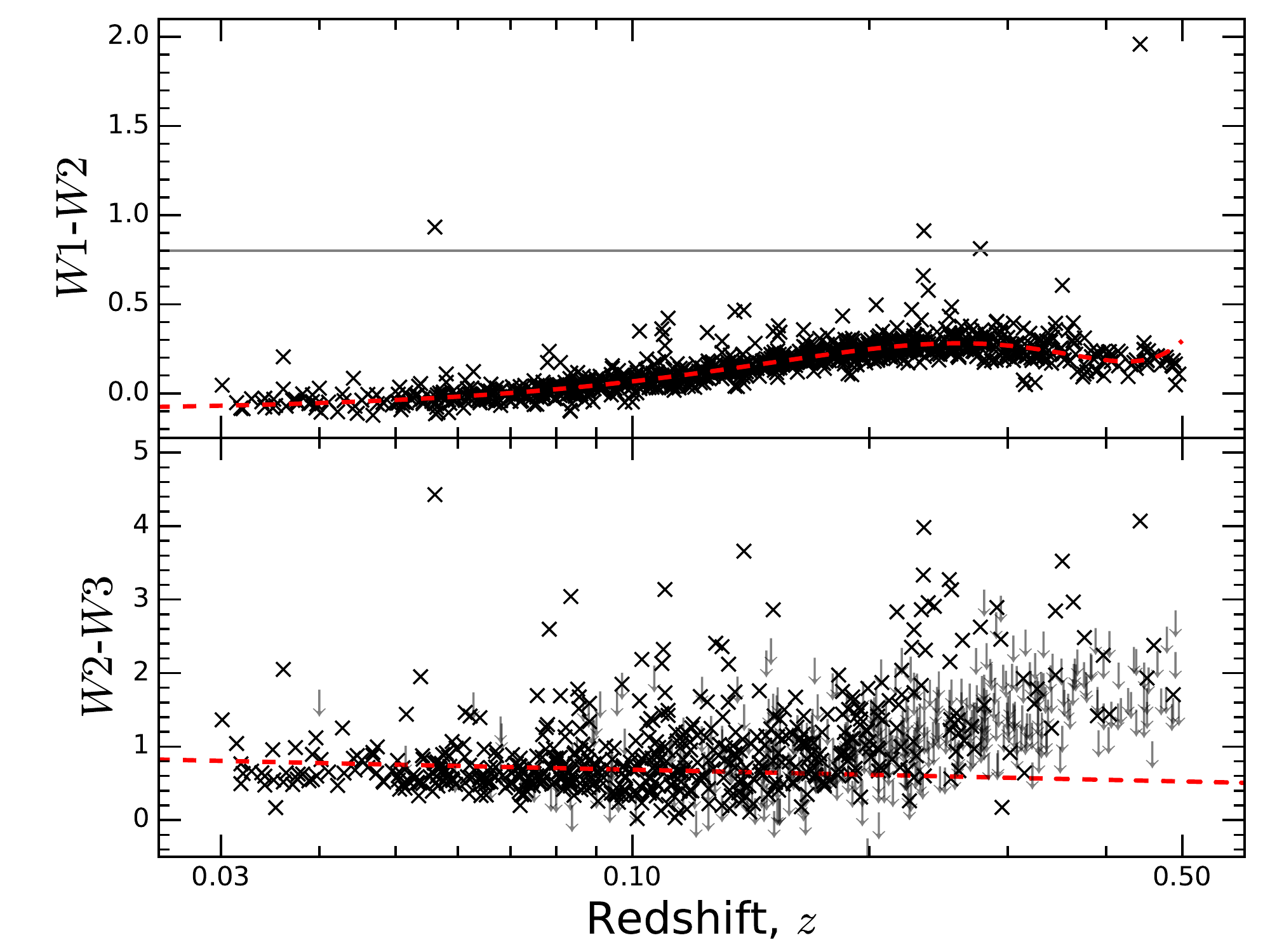}}
 \caption{The \wb1-\wb2 ($3.4-4.6\micron$) and \wb2-\wb3 ($4.6-12\micron$) colour evolution of the BCGs, with respect to redshift. The grey arrows indicate sources where \wb3 $S/N<$\SNfit and the red dashed lines indicate the assumed best fit for a passively evolving galaxy (see main text). The solid grey line in the top panel indicates the selection criteria of \wc$>0.8$ mags for strong AGN. 
}
 \label{Fig:w12-z}
\end{figure}

\begin{figure}
\centering
 \resizebox{\columnwidth}{!}
 {\includegraphics{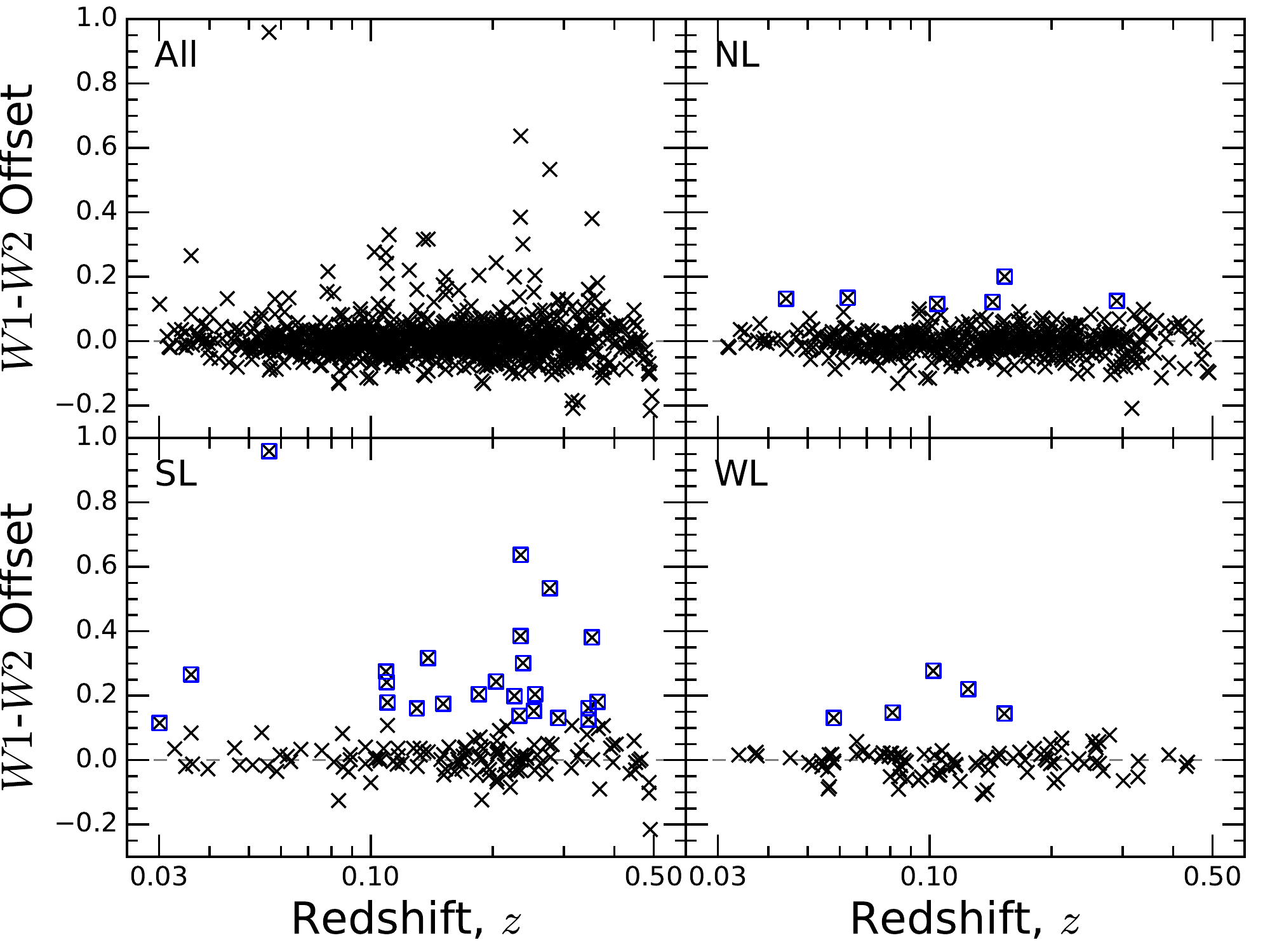}}
 \caption{The \wise \wc  ($3.4-4.6\micron$) BCG colour offset against redshift. \SubCapt Note there is another point with a \wc offset of 1.78, corresponding to MACSJ0913.7+4056, which is omitted here for the sake of visual clarity. }
 \label{Fig:w12off-z}
\end{figure}

\begin{figure}
\centering
 \resizebox{\columnwidth}{!}
 {\includegraphics{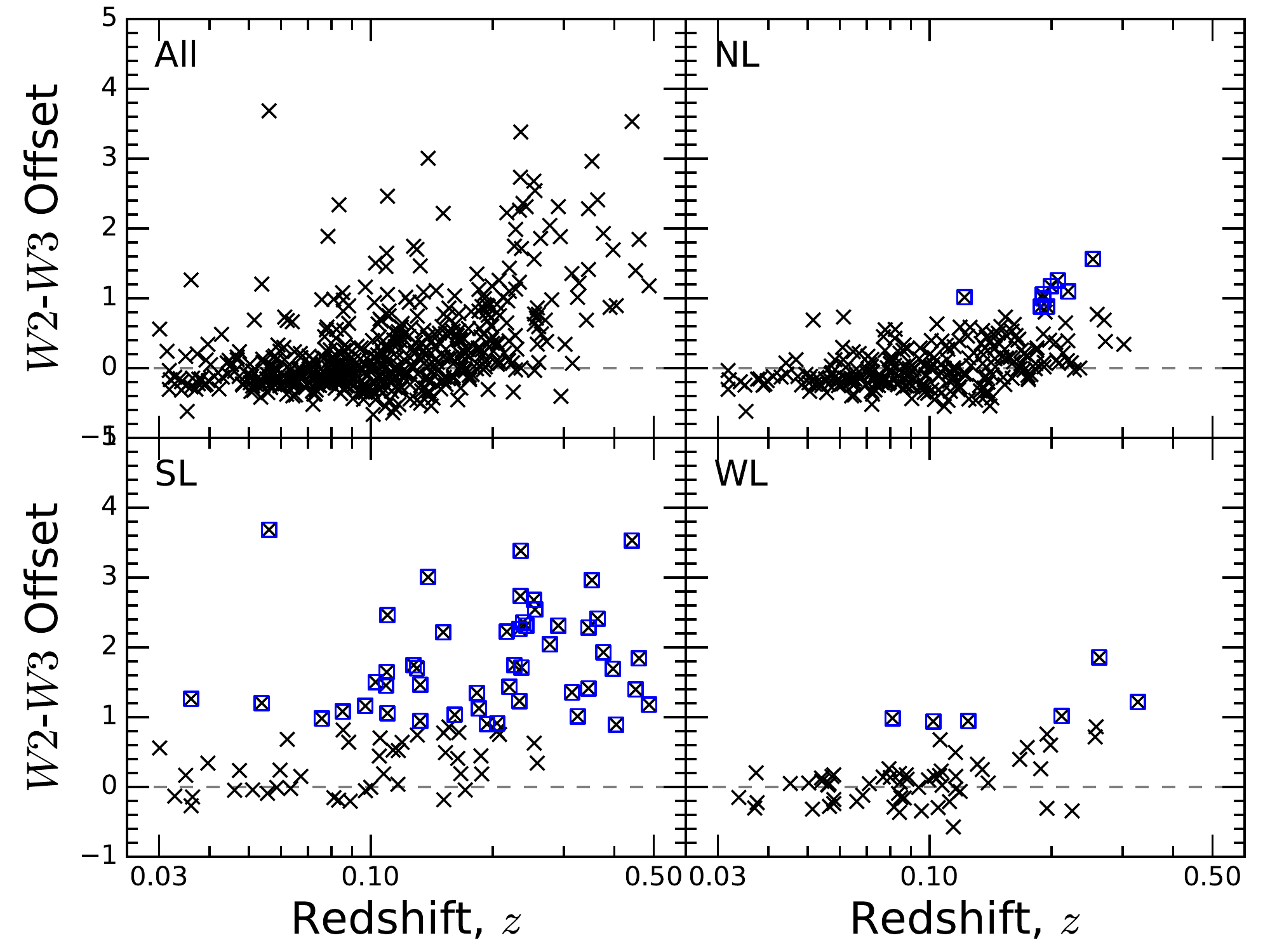}}
 \caption{The \wise \wcol ($4.6-12\micron$) colour offset against redshift for BCGs with a \wb3 S/N$>$\SNfit (\WISESNnum BCGs). \SubCapt}
 \label{Fig:w23Off-z}
\end{figure}

\begin{figure}
\centering
 \resizebox{\columnwidth}{!}
 {\includegraphics{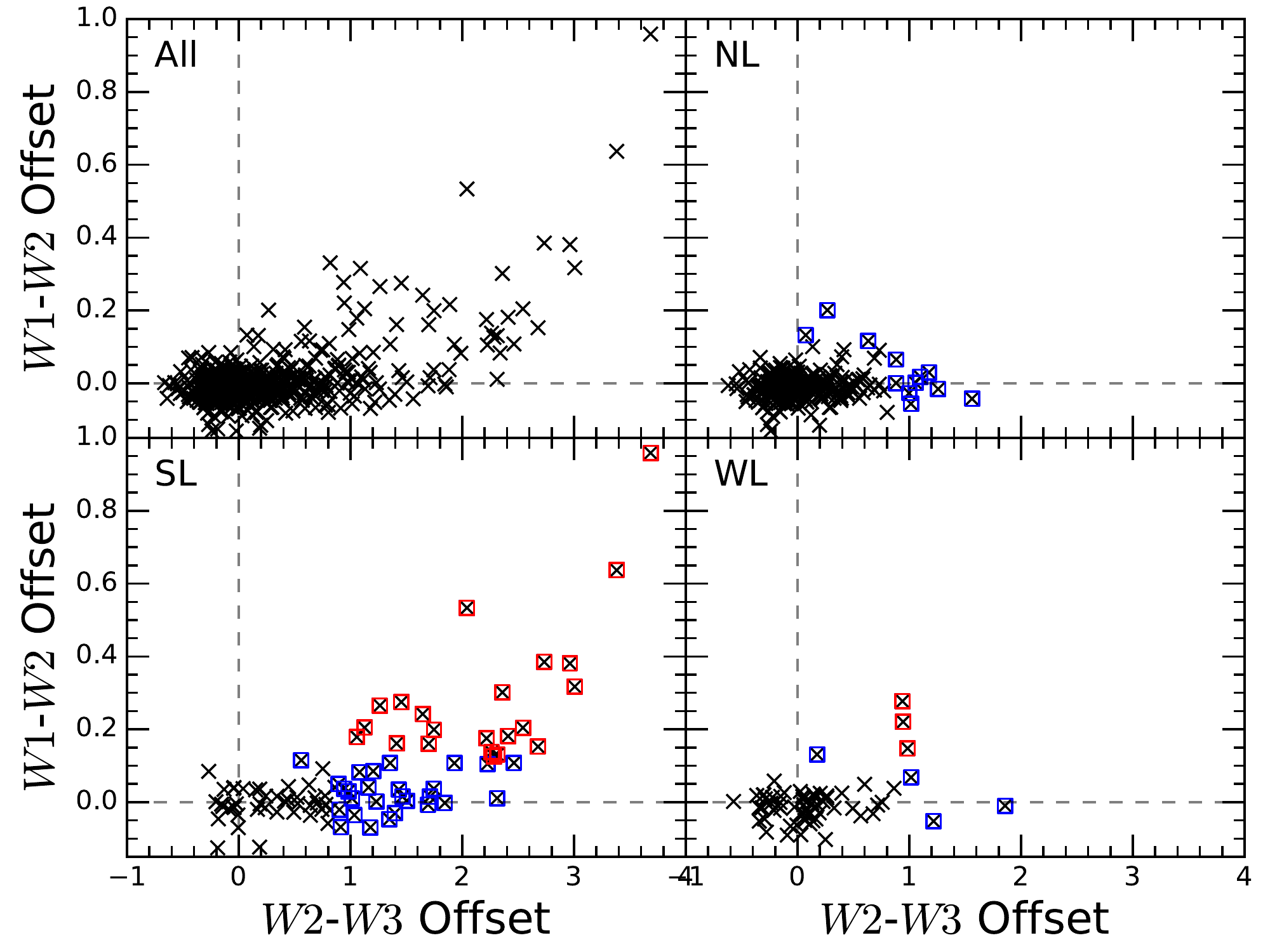}}
 \caption{The \wc offset against \wcol offset for BCGs. \SubCapt The red squares correspond to galaxies significantly offset in both colours and thus likely indicate BCGs hosting AGN. Note there is another point with \wc and \wcol offsets of 1.78 and 3.52 respectively, corresponding to MACSJ0913.7+4056, which is omitted here for the sake of visual clarity.}
 \label{Fig:w31-w21}
\end{figure}

\subsection{Mid-IR Analysis and Results}
\subsubsection{BCG Luminosity}
\indent In  Fig. \ref{Fig:w1Lx}, we find that the BCG \wise \wb1--band luminosity correlates with the X-ray luminosity of the host cluster. This reinforces the results from our optical analysis (Fig. \ref{fig:Lx_BCGi_s12}), but with less scatter in the Mid-IR. The interpretation is that BCGs with higher stellar mass tend to reside in more massive clusters, as suggested from Near-IR observations also (e.g. \citealt{Lin+04, Stott+08, Stott+10,  Lidman+12}). Applying a linear regression BCES bisector fit \citep{Isobe+90} we find, \begin{equation} L_{W1} = (-1.61\pm0.02) \mathrm{log}_{10}(E(z)^{-1}L_{\mathrm{X}})-(2.57\pm0.02), \end{equation} where $L_{W1}$ is the BCG \wb1-band luminosity (mags), \Lx is the X-ray luminosity ($10^{44}\,\mathrm{erg\,s}^{-1}$) and $E(z) = [\Omega_m(1+z)^3+\Omega_{\Lambda}]^{1/2}$. 
\subsubsection{BCG Colours}

\indent In the IR the contribution from stellar mass decreases with wavelength, becoming more strongly dependent on dust emission at larger wavelengths. Hence in the Mid-IR, particularly at 12\micron, we are sensitive to reprocessed emission from dust. An active BCG, with recent star formation and/or AGN activity, is consequently expected to show an excess in Mid-IR emission as the dust gets heated by the hot young stars and/or AGN. The colour of an active galaxy would as a result be redder than that of a passive galaxy, because the dust contribution is more significant in the redder band. As AGN output can heat dust to $T>80K$, the most extreme cases of Mid-IR excess are likely due to AGN contribution. 

\indent All BCGs are well detected in \wb1 and \wb2, but only \WISESNnum BCGs are robustly detected (i.e. $S/N>\SNfit$) in \wb3, due to the declining continuum in the SEDs of passive galaxies with respect to wavelength in the IR. In Fig. \ref{Fig:flux-z} we show the \wb1, \wb2 and \wb3 flux against redshift, with arrows indicating upper limits (i.e. where $S/N<\SNfit$). The proportion of BCGs detected in \wb3 is clearly a function of redshift and beyond $z\sim0.15$ the quiescent BCG population is not well sampled. Fortunately BCGs with a clear \wb3 excess continue to have robust detections across the redshift range and hence we expect this to have minimal effect on the detection of active BCGs. In order to define a \wb3 excess however this does require an extrapolation of the \wb3 flux-redshift relation for the bulk population of passive BCGs. Firstly a best fit is made to the bulk population, (ignoring those with clear excess), for \wb1 and \wb2. From the Figure we see the relation between these two is very similar and the relation for \wb3 is expected to be similar also. Hence using the \wb2-$z$ relation as a template, we iteratively determine the best fit to those BCGs with $S/N>\SNfit$ and $z<0.15$, again ignoring the BCGs with a clear \wb3 excess, and extrapolate this relation to the higher $z$, where our bulk quiescent population is lost to the flux limit. 

\indent We can also deduce from Fig. \ref{Fig:flux-z} that \wise can measure BCG fluxes beyond $z>0.5$. In \wb1 and \wb2 it appears we can continue detecting the bulk population further, but crucially for studies of activity, we expect to be able to measure excesses even further. This will be a useful factor when comparing BCG properties of the low-$z$ Universe to those at higher $z$. For instance, the most extreme systems at $z>0.5$, such as the Phoenix cluster ($z=0.598$, \citealt{McDonald+12}), that is an AGN-dominated BCG with \wb3 magnitude of 7.93~mag, would be detectable with {\it WISE} to at least $z=1.5$. So, while the less active AGN are lost at $z>0.3$, we can still identify the most active systems in any sample, such as the XCS \citep{Romer+01} in the X-ray, or MaDCoWS \citep{Stanford+14} in the Mid-IR.

\indent In order to test for a Mid-IR excess we explore the Mid-IR colours, \wc and \wcol. In Fig. \ref{Fig:w12-z} the \wc ($3.6-4.6\micron$) colour against redshift shows that the passive BCG population undergoes a peaked evolution with redshift - due to the redshift of a continuum break in the SED here. This effect has sometimes been ignored in the literature, where conclusions are drawn from raw colours only (e.g. \citealt{Fraser-McKelvie+15, Quillen+08}). Fig. \ref{Fig:w12-z} illustrates that if using just raw colours, \wb2 excesses at redshifts below the peak are lost to the redshift relation and hence some star forming galaxies would be missed. This illustrates why one should consider a colour offset (as used here) or a normalised flux ratio (such as in \citealt{Hoffer+12}). Having already established most BCGs are indeed passive we can hence determine a \wc colour offset by collapsing along the best fit line to the bulk passive trend in Fig. \ref{Fig:w12-z}, with the offset defined as the difference in the measured colour and that predicted by this best fit. 

\indent The \wc offset shown in Fig. \ref{Fig:w12off-z} show that there are a number of BCGs which exhibit a significant offset and that these correlate with the optical emission line status. For the non-line emitting BCGs, i.e. passive BCGs, we measure a scatter with a standard deviation, $\sigma_{p} = 0.04$ mags and a mean of zero. If we then define a \wb2 excess as an \wc offset $>2.5\sigma_{p}$, we find $5\%$ of the total show an excess, increasing to $17\%$ for the subsample of strong line emitters. 

\indent Enhanced \wb2 emission is often an indicator of AGN activity in a galaxy  (resulting from the nature of the power spectrum of AGN \citep{Odea+08}). Using the basic AGN selection criteria of \wc$>0.8$ from \cite{Assef+13} we recover four BCGs hosting a strong AGN. These four are the BCGs of Zw 2089 \citep{Russell+13}, Cygnus-A \citep{Russell+13}, MACSJ0913.7+4056 \citep{Hlavacek+13}, and PKS2338+000. 
This allows us to determine a conservative lower limit of four BCGs hosting strong AGN in our sample (<1\%). However the reliability and completeness of a simple colour AGN selection method assumes highly luminous AGN only, and are calibrated on `typical' AGN host galaxies. BCGs are anything but typical and hence it is inevitable that the host BCG has a significant contribution to the \wb1 flux, diluting the relative \wb2 excess. Hence we are almost certainly underestimating the overall AGN fraction with other AGN host galaxies just falling short of this formal selection cut. Nonetheless, this result shows that whilst ongoing strong AGN are extremely rare, the AGN duty cycle is non-zero in these galaxies.

\indent Fig. \ref{Fig:w23Off-z} shows the \wcol ($4.6-12\micron$) colour offset. This is difference between the measured \wcol and the estimated best fit for \wcol, given by the \wb2 best fit minus the expected \wb3 best fit. Once again we see that non-line emitting BCGs do not generally show any Mid-IR excess, with a mean offset of zero, consistent with passivity. With a scatter of  $\sigma_{p} = 0.35$ mags in the non-line emitters we find at least $8\%$ of total BCGs show a significant colour offset (i.e. an offset $>2.5\sigma_{p}$). 
This fraction is increased to $35\%$ for the strong line emitting BCGs. As expected weak line emitters behave somewhere in between, with most near zero but with a handful showing a significant offset ($6\%$). The reader should note that the apparent deficit of non-offset galaxies at high redshift is a result of the passive BCG population lacking robust detections in \wb3 at $z\gtrsim0.15$. One should also note that the fraction of BCGs said to show a significant offset are evaluated as a proportion of the total sample, under the assumption that the undetected BCGs are exclusively passive. This means that the fractions are in fact lower limits, since although most BCGs with Mid-IR excess are still detected, there could be some BCGs with a modest, but still significant offset, where robust detections are not possible in \wb3.
\\ \indent In Fig. \ref{Fig:w31-w21} we plot the \wc and \wcol colour offsets against each other for BCGs with $S/N>\SNfit$. There is a clearly a correlation between a \wc offset and \wcol offset, with the most extreme offsets in one corresponding to the most extreme offset in the other. As discussed above, the BCGs with the largest offsets in both the Mid-IR colours are likely to be a result of AGN contribution within the BCG. Whereas the BCGs with a significant offset in \wcol, but which are do not show a significant \wb2 excess are likely to be dominated by star formation \citep{Donahue+11}. Overall, we find at least $9\%$ of our BCGs show an excess in either \wb2 or \wb3, increasing to $36\%$ for the strong line emitting subsample (see Table \ref{tab:fract}).
%

\begin{figure}
\begin{minipage}{\textwidth}
\includegraphics[width = 0.475\textwidth]{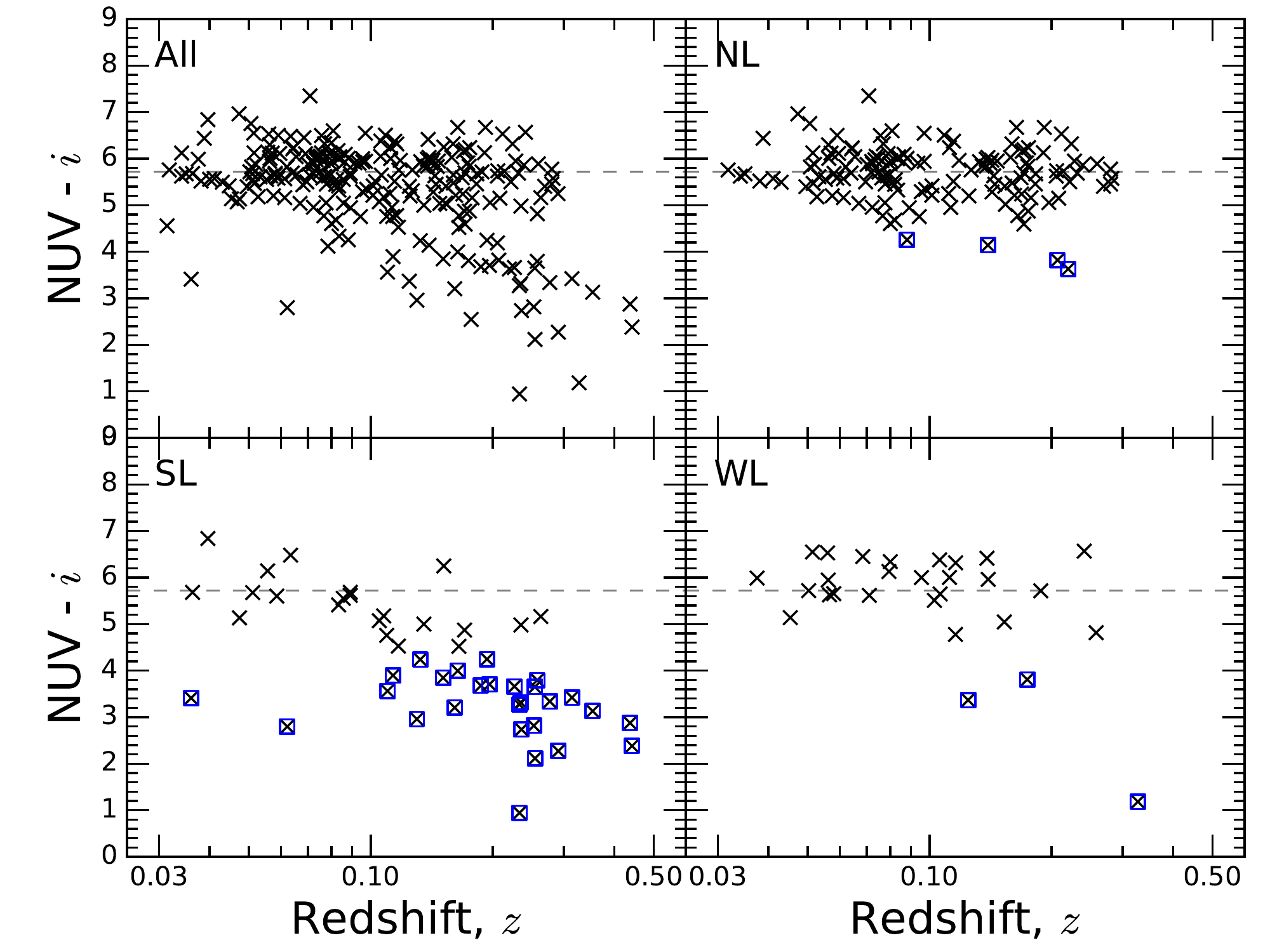}
\end{minipage}
 \caption{NUV--\ib against redshift for the \NUVrob BCGs detected in the NUV with $S/N$>\SNfit. The dashed line indicates the median colour of the passive, non-line emitting, BCGs at 5.80 mags. We define subsequent NUV--\ib colour offsets relative to this line. \SubCapt}
 \label{fig:UV}
\end{figure}

\subsection{UV Analysis and Results}
\indent In the UV wavelength regime we are particularly sensitive to emission from young O and B type stars, hence recent star formation is characterised by an enhanced UV emission. We investigate the BCG NUV--\ib colour against redshift, (Fig. \ref{fig:UV}), in order to test for enhanced UV emission in our BCG sample. We use the \ib band photometry because at these redshifts it is unaffected by the shift of the $4000$\r{A} break. There appears to be no significant redshift evolution in the NUV--\ib colour for the quiescent BCGs and so we define a UV offset as the difference between the measured NUV--$i$ and the median NUV--$i$ value of $5.80$. The non-line emitting BCGs form a relatively tight cloud around this median, with a standard deviation of $\sigma_{p}=0.57$ mags, consistent with passivity. 

As a fraction of the \NUVrob BCGs with robust NUV detections, we find $16\%$, $32\%$ and $4\%$ show a significant offset, (defined as an NUV--$i$ offset $>2.5\sigma_p$), in the full, strong line and weak line emitting samples respectively. However we stress that the UV detections are not complete in sky coverage, or of uniform depth for those observed. Because of this we preferentially select galaxies with a UV excess. Hence we are only able to determine lower limits, as a fraction of the total BCG sample. With this in mind, we find that at least $4\%$ of BCGs show a UV excess. This fraction is increased to $19\%$ of BCGs known to exhibit strong optical emission lines and is $3\%$ for those with weak lines. (As summarised in Table \ref{tab:fract}).
  

\subsection{Comparing the Optical, UV and Mid-IR Photometry.}
\begin{figure*}
\begin{minipage}{\textwidth}
{\includegraphics[width = 0.475\textwidth]{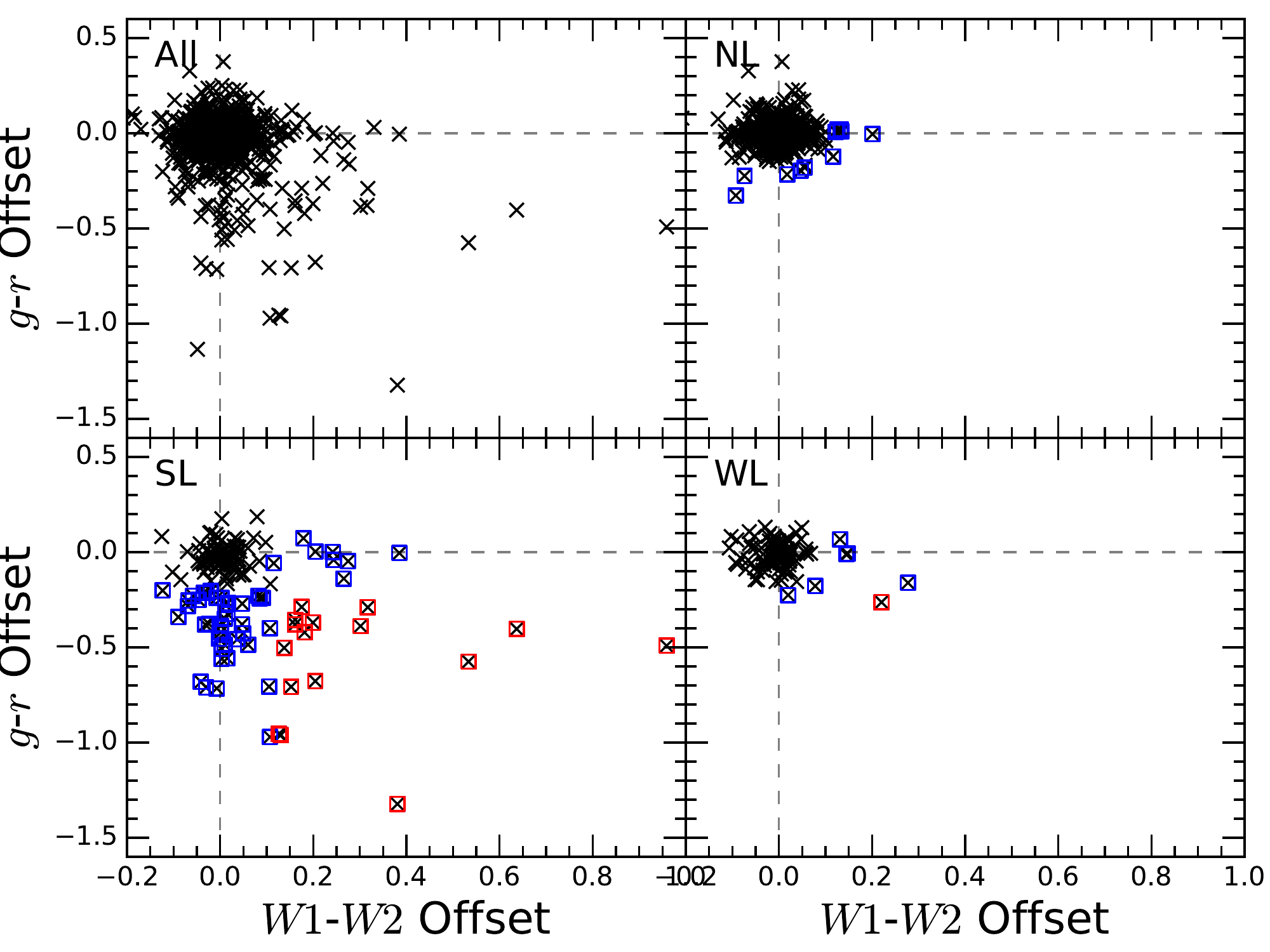}}
\hfill
{\includegraphics[width = 0.475\textwidth]{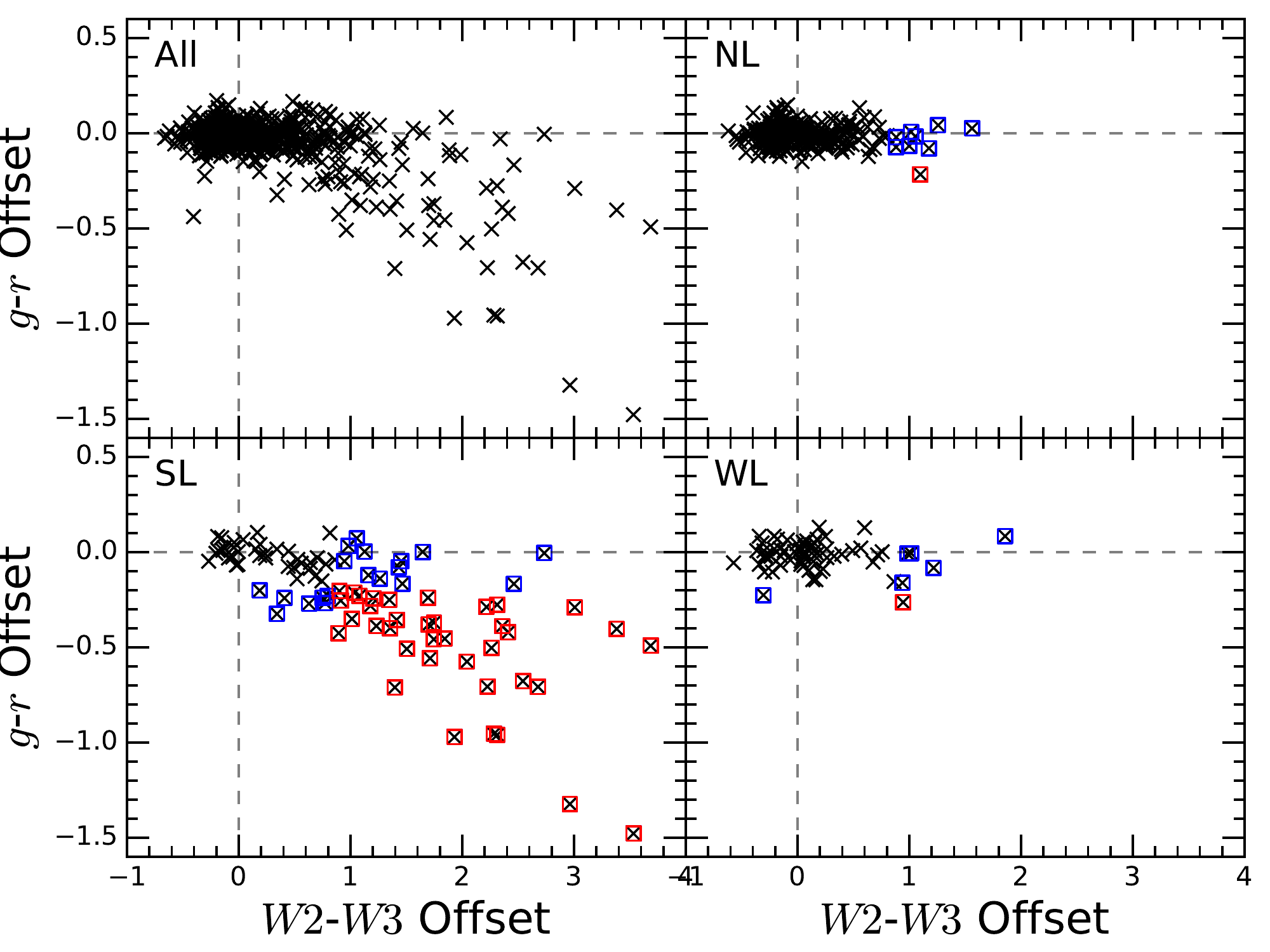}}
\vspace{0.7cm}
{\includegraphics[width = 0.475\textwidth]{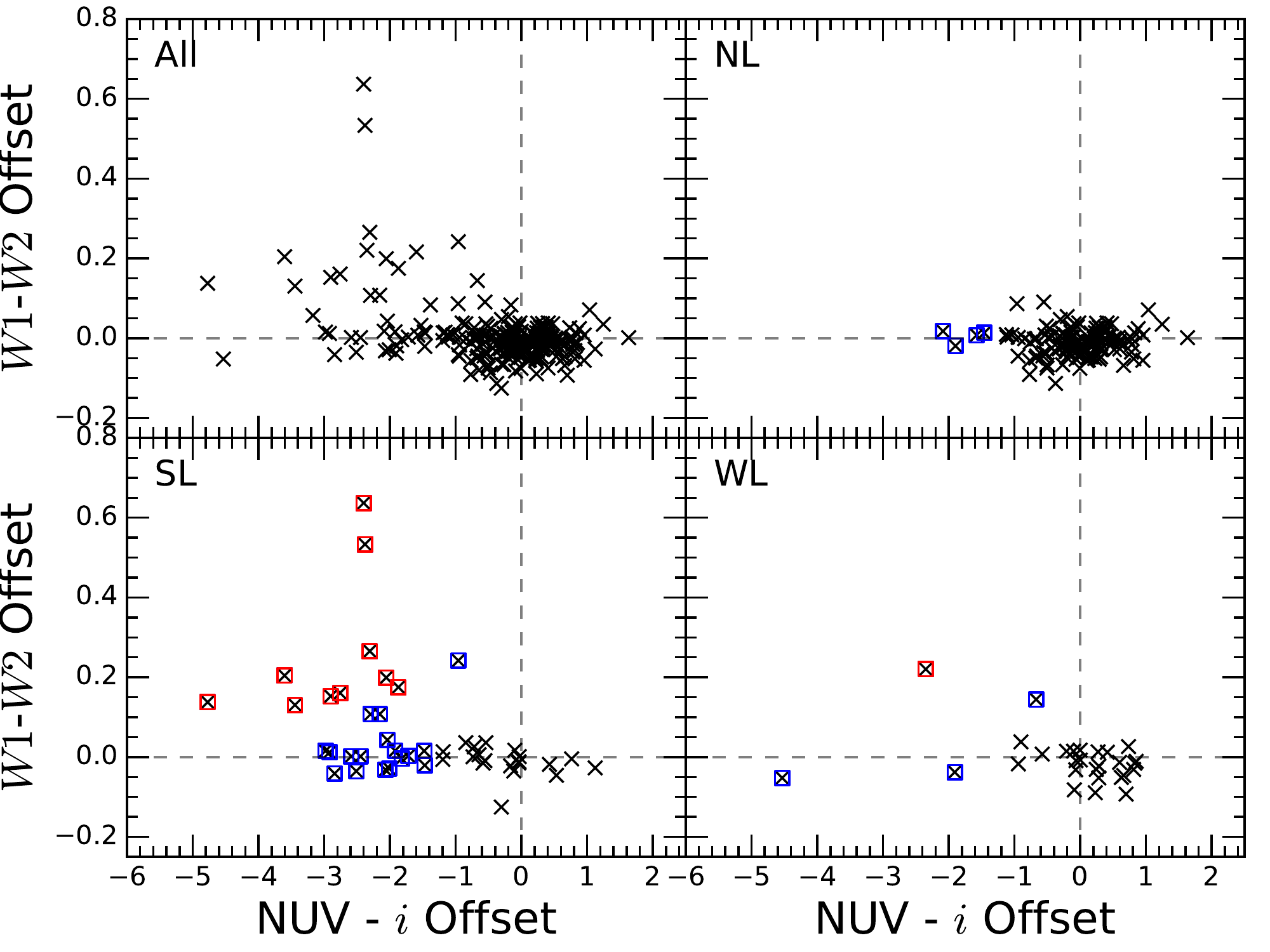}}
\hfill
{\includegraphics[width = 0.475\textwidth]{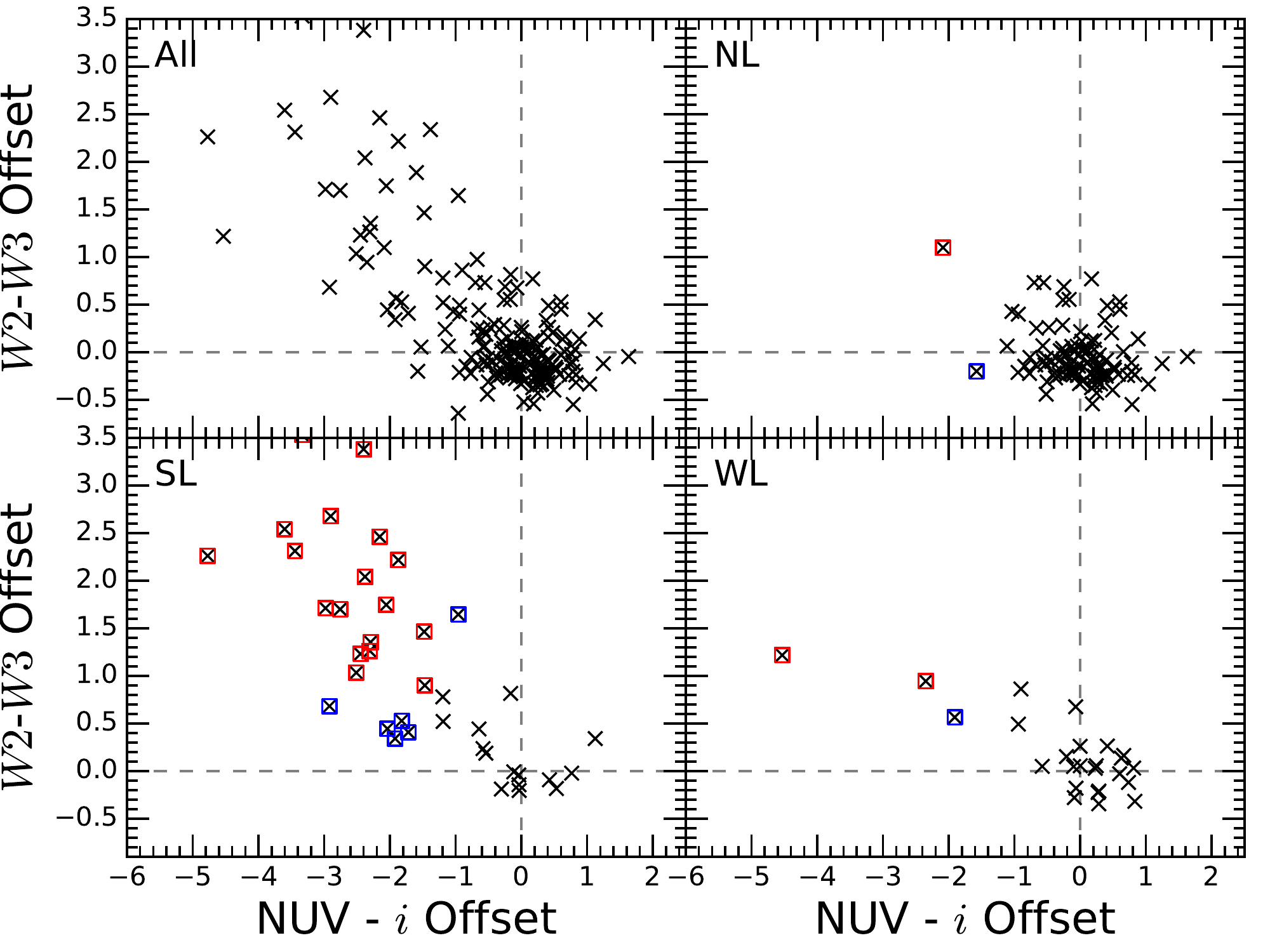}}
\vspace{0.7cm}
\centering
 {\includegraphics[width = 0.475\textwidth]{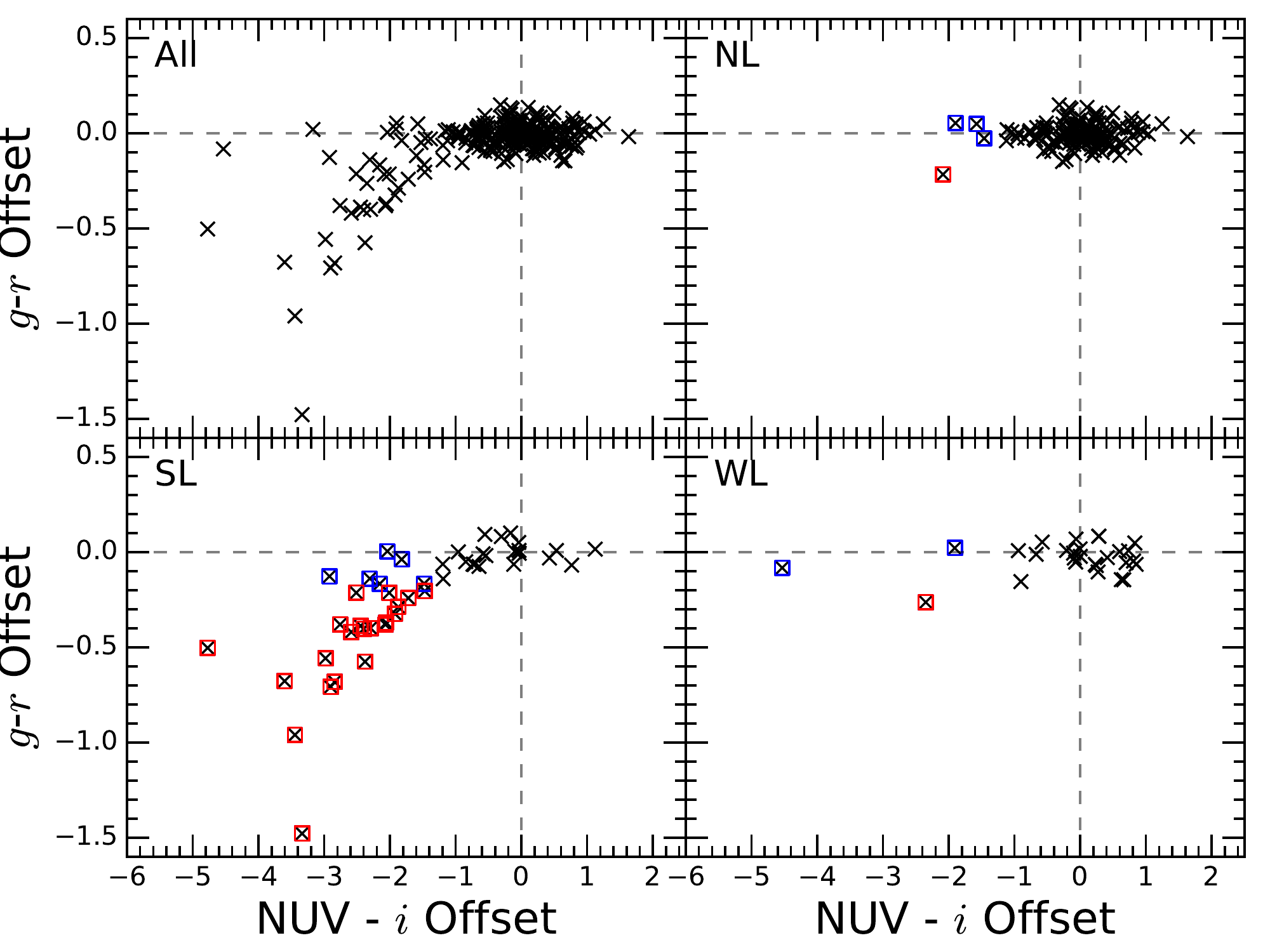}}
\end{minipage}
 \caption{Plots of the various colour offsets against one another. The panels are split into: full sample (top left), non line emitters (top right), strong line emitters (bottom left) and weak line emitters (bottom right.) The blue squares indicate those BCGs with colour offsets in excess of $>2.5\sigma_p$ from zero in one colour and the red squares indicate those BCGs with significant colour offsets in both colours. Note that we only display sources with a $S/N>$\SNfit, which limits the number of \wb3 and NUV detections.}
    \label{Fig:coloff-coloff}

\end{figure*}

In Fig. \ref{Fig:coloff-coloff} we display the PS1, \wise and \galex colour offsets against one another. We see that BCGs with a significant colour offset at one wavelength are likely to, but not necessarily to, exhibit a significant offset in another. This tells us that the colours at both wavelengths are likely to be due to the same phenomena. Fig. \ref{Fig:coloff-coloff} also illustrates the importance of using multi-wavelength photometry in such an analysis. We see that not all colour bands measure BCG activity equally, due to the complimentary nature of different wavelength observations. Some bands are more sensitive to certain phenomena than others, and over different time-scales. In the UV for example we are especially sensitive to ongoing, unobscured star formation, whereas in the Mid-IR we can measure obscured star formation/AGN activity, and over longer time-scales. 

\indent In Fig.~\ref{Fig:coloff-coloff} we see only one non-line emitting BCG, that of A1704, which has a significant offset in more than one colour index, (with a significant offset in \grb, NUV--\ib and \wcol). This particular galaxy has an SDSS spectra of its core, which shows no emission lines. However, the imaging clearly shows an optically blue component within the BCG, but offset from the centre of the galaxy. Our prediction is that if the spectral fibre had been placed on this region then it would likely show strong emission lines, (however projection cannot be ruled out at this stage). Optical line emission offset from the BCG has been observed in a number of cluster cores (\citealt{Hamer+12}; Green at al. in prep) and hence is something to take into consideration when analysing activity in cluster cores. 

\indent The lack of non-line emitters, with significant offsets in multiple colours, supports our assumption that these are a fair representation of passively evolving galaxies. We believe that the few non-line emitters which are classed as active in our analysis are a result of scatter about a fixed cut-off value. The fraction of active non-line emitting BCGs can consequently provide a rough measure of the uncertainty in the active BCG fractions reported. The scatter about this colour cut-off can result from a number of sources, but primarily photometric uncertainty and contamination. A known source of contamination, in at least two of the BCGs in this sample, is from lensed galaxies. There are clear gravitational arcs very close to the BCGs of A521 \citep{Richard+10} and A2104 \citep{Pierre+94}. Both clusters have archival \textit{Spitzer} observations, in which the arcs are clearly luminous in the Mid-IR. Unfortunately the lower resolution of \wise means the emission of the BCGs and arcs are blended, probably accounting for the significant offset seen in the \wcol and \wc for A521 and A2104 respectively. (Note: an excess of $24\micron$ emission was previously noted for the BCG of A521 in \cite{Hoffer+12}, which is also likely a result of source blending of the arc). Given the rarity of gravitational arcs with low radius of curvature, this is unlikely to be a significant source of scatter. However, a few more may exist in such a large cluster sample. 

\subsection{BCG Activity and Cluster X-ray Luminosity}
\begin{figure*}
\centering
 \resizebox{1.25\columnwidth}{!}
 {\includegraphics{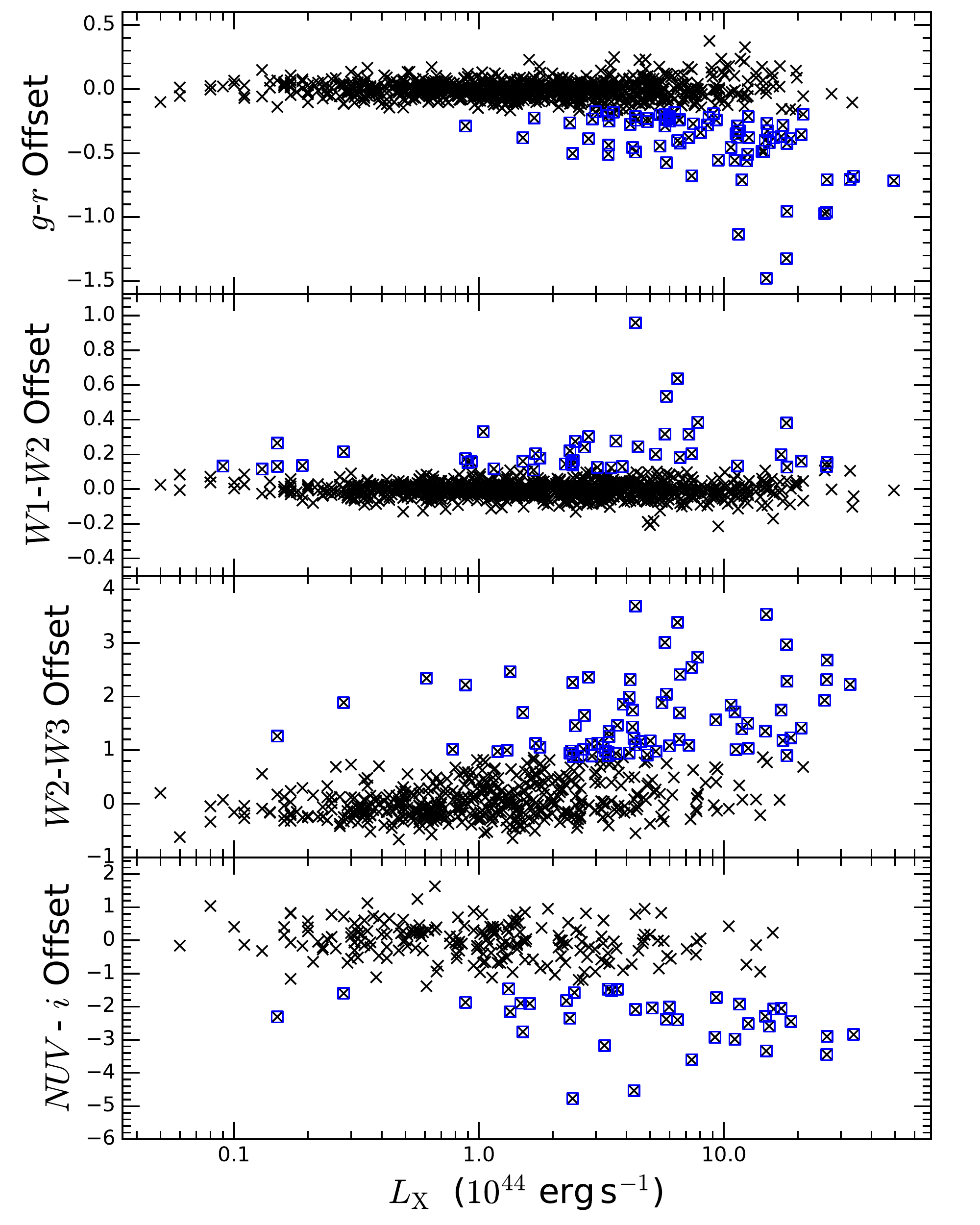}}
 \caption{BCG colour offsets, against the X-ray luminosity of the host cluster, in the optical (top panel), Mid-IR (second and third panels) and UV (bottom panel). We see that BCGs with colours which deviate from the quiescent BCG population tend to belong to more X-ray luminous clusters, consistent with BCG activity having a strong environmental dependence. Note that the first two panels are complete, but that the \wb3 and NUV detections are limited by $S/N$. So, since $S/N$ is systematically lower at high redshift, high $L_{\mathrm{X}}$ clusters are preferentially excluded by selection effects.}
\label{Fig:LxRelations}
\end{figure*}

The X-ray luminosity of a cluster is a combination of both the peaked emission of the core and the extended cluster emission. Hence it is reasonable to assume that, given the observed strong correlation between cool core clusters and BCG activity (see Introduction), any relationship seen between tracers of activity and \Lx is indicative of this underlying link and suggestive that these clusters are also cool cores. In Fig.~\ref{Fig:LxRelations} we plot colour offsets against cluster \Lx and find that BCGs with significant optical, Mid-IR and UV colour offsets tend to belong to more X-ray luminous clusters. However, a high X-ray luminosity does not necessarily imply BCG activity. The number and proportion of BCGs with significant colour offsets, in bins of X-ray luminosity, are presented in Table \ref{Table:Lx}, which confirms there is an apparent association between BCG activity and host cluster gas properties. This result agrees with the observations of \cite{Odea+08} who found BCG IR excesses preferentially in higher X-ray luminosity clusters.

\begin{figure}
\centering
 \resizebox{\columnwidth}{!}
 {\includegraphics{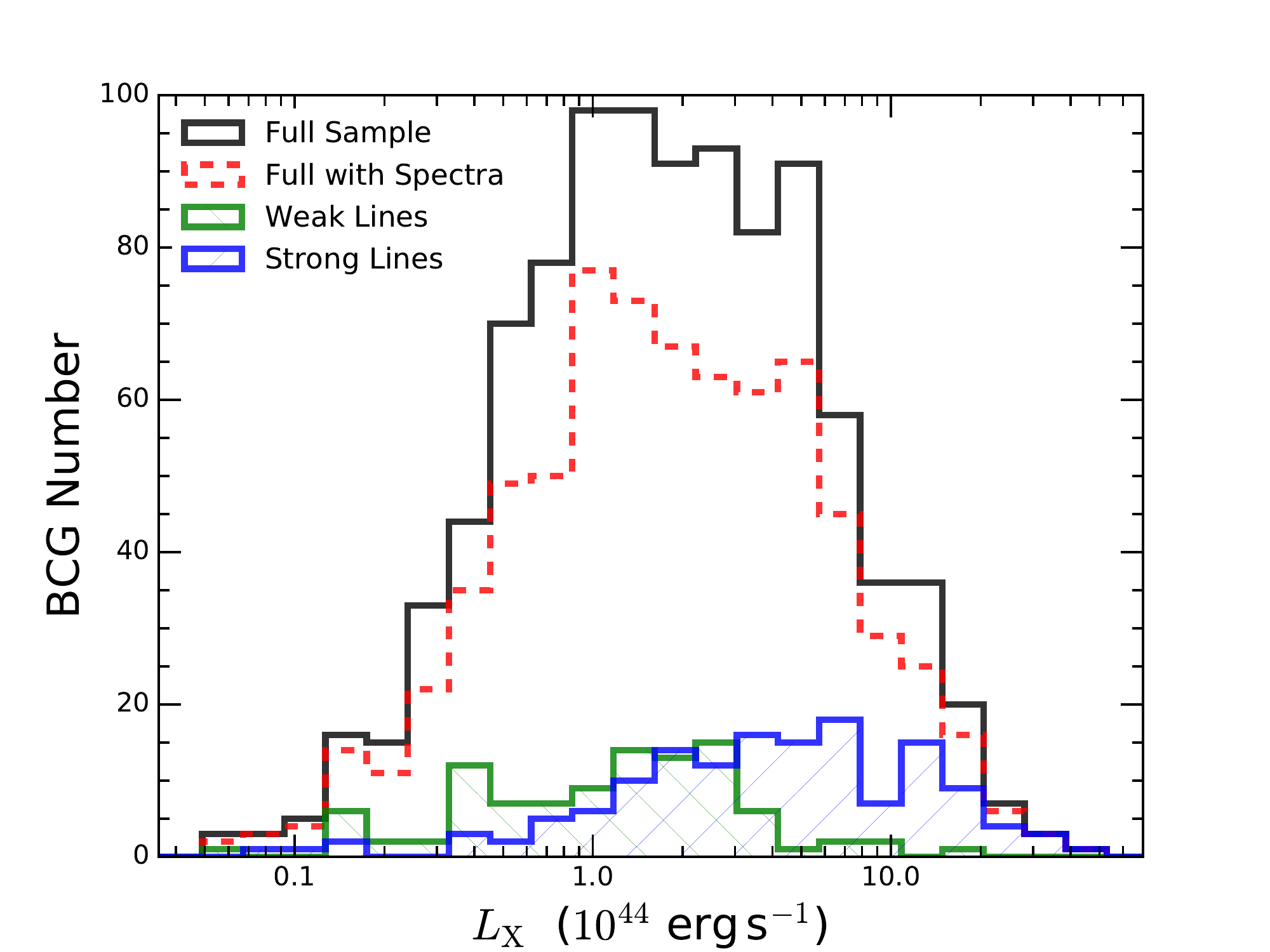}}
 \caption{A histogram showing the host cluster X-ray luminosity distribution for the full sample of BCGs (solid black), the full sample of BCGs with spectroscopy (red dotted), as well as subsamples of BCGs with strong (blue hatch) and weak (green hatched) emission lines. We see that the presence of strong lines (hence BCG activity) is preferentially in the most X-ray luminous clusters. }
\label{Fig:LinesLx}
\end{figure}

 \indent The histograms in Fig.~\ref{Fig:LinesLx} show the X-ray distributions for clusters with line emitting BCGs and how this compares with the overall distribution. The presence of optical emission lines is strongly associated with the X-ray properties of the host clusters, with strong line emitters preferentially located in the more X-ray luminous clusters. From Table \ref{Table:Lx}, we see that the fraction of BCGs which show signs of activity through the presence of strong lines increases with respect to X-ray luminosity. This positive trend is in agreement with the results of \cite{Samuele+11} who found fractions of $30\%$, $18\%$ and $16\%$ for the \cite{Crawford+99} BCS sample of BCGs for the same \Lx ranges respectively. Note that the apparent deficit in weak line emitting BCGs at higher X-ray luminosities is likely to be an observational effect, in which high \Lx clusters are more likely to be high redshift, where the detection of weak lines is more difficult.

\indent The presence of emission lines in BCGs can be used as tracer of cool core clusters (\citealt{Heckman+81, Hu+85, Johnstone+87, Heckman+89, Crawford+99, McDonald+11}). So given we find a link between photometric colours and emission line status, we can then deduce that peculiar photometric colours also acts as a tracer of cool cores. This reinforces the assumptions given above that the peculiar colour and \Lx association is reflective of the likely cool core status of these particular clusters.

\indent The strong association between BCG activity and total X-ray luminosity of the host cluster, evident through the photometric excess and optical line emission, has important implications regarding cluster selection effects. Specifically, the fraction of active BCGs one discovers in any given cluster sample is dependent on the X-ray luminosity distribution of that sample. For example, \cite{Stott+12} do not find any unusual coloured BCGs in \grb in the XCS sample, but this is not surprising given the low X-ray luminosity nature of that particular cluster sample. With this in mind we investigate the active BCG fraction for volume complete subsamples and make comparisons to the literature. 

\indent If we consider an ``X-ray luminosity complete'' cluster subsample defined by \Lx$\geq1\times10^{44}\,\mathrm{erg\,s}^{-1}$ and $z\leq0.2$, (as indicated in Figure \ref{fig:Lx_z}), we find active BCG fractions as presented in Table \ref{Table:completefractions}, specifically that at least $14\%$ of clusters deviate from passivity in at least one colour. This is comparable to our overall figure of $14\%$ because the lower \Lx and higher $z$ sections of our full sample cancel out their relative effects.
\\ \indent Similarly, in Table~\ref{Table:FM+15Fractions} we summarise the BCG fractions within the subsample defined by \Lx$\geq1\times10^{44}\,\mathrm{erg\,s}^{-1}$ and $z\leq0.1$. This \Lx--$z$ distribution was chosen to match that of \cite{Fraser-McKelvie+15}, who conclude that $<1\%$ of their BCGs are star forming using \wise. We conversely find that $>8\%$ of our BCGs in the same \Lx--$z$ space have a colour offset in \wise, increasing to at least $10\%$ if you include the optical and UV data. This disagreement we think reflects their use of raw \wise colours and of a simple colour cut to select star forming galaxies. In their plot a significant number of BCGs are offset from the bulk passive cloud in colour--colour space, yet do not satisfy the colour cut used. One would expect very few BCGs in the Universe to be forming stars on the level one would expect for a classically ``star forming'' galaxy (like a spiral), but instead one should consider how a modest amount of star formation would make the \wise colours differ from the bulk passive cloud of BCGs in colour space.  

\indent An additional observational consideration for BCG activity is that the peaked X-ray emission in cool core clusters could be misinterpreted as a point source in shallow X-ray data. If the BCG of the cluster shows strong high ionisation line emission, usually, and understandably, that X-ray source is identified as an AGN. But, the cluster may still be a significant contributor. This is best illustrated by the Phoenix cluster, \citep{McDonald+12}, where the most X-ray luminous cluster was only identified because of the SPT SZ effect detection. Therefore caution is needed when interpreting the perceived rarity of BCGs hosting strong AGN as this identification bias could be masking systems such as H1821+644 \citep{Russell+10} and 3C186 \citep{Siemiginowska+10} where the cluster X-ray emission is a minority contributor to the total flux, but it would still be above the X-ray flux limit of a survey once the AGN contribution is removed. 

\begin{table}
 \centering
 \caption{Percentage of BCGs with an optical (Opt.), Mid-IR or UV colour offset $>2.5\sigma_p$, where $\sigma_p$ is the scatter for the non-line emitting (passive) BCGs. The percentage of BCGs with a significant offset in any one of the colours is given in the ``Combined'' column. The ``Radio'' column indicates a detection only, not an excess. Percentages are presented as fractions of the full sample (All) and strong line emitting (SL), weak line emitting (WL) and  non-line emitting (NL) subsamples. }
 \begin{tabular}{lcccccc}
  \hline
  \hline
    & BCGs & Opt. & MIR & UV & Combined & Radio\\
    & (\#) & (\%) & (\%) & (\%) & (\%) & (\%)\\
  \hline
All & 981 & 8 & 9 & 4 & 14  & 52\\
SL & 144 & 39 & 35 & 19 & 51 & 84 \\
WL & 100 & 3 & 8 & 3 & 11 & 43 \\
NL & 476 & 1 & 3 & 1 & 5  & 43\\
\hline
\label{tab:fract}
\end{tabular}
\end{table}

\begin{table}
 \centering
 \caption{Number of BCGs with significant colour offsets, and optical emission lines, in bins of cluster X-ray luminosity. The results suggest active BCGs are preferentially found in high \Lx clusters. (Note the apparent decline in the weak line fraction in the highest \Lx is likely a redshift dependence effect - specifically high \Lx clusters are preferentially at high $z$, where the $S/N$ in observations makes identifying weak lines difficult.)}
 \begin{tabular}{ccccc}
  \hline
  \hline
    & Total & \Lx$\geq45$ & \Lx$\geq44$ & $42<$\Lx$<44$\\
   & 981 & 76 & 671 & 310\\
  \hline
\grb & 79 ($8\%$) & 35 ($46\%$) & 78 ($12\%$) & 1 ($1\%$)\\
\wc & 45 ($5\%$) & 8 ($11\%$) & 36 ($5\%$) & 9 ($3\%$)\\
\wcol & 75 ($8\%$) & 19 ($25\%$) & 70 ($10\%$) & 5 ($2\%$)\\
NUV-$i$ & 37 ($4\%$) & 12 ($16\%$) & 34 ($5\%$) & 3 ($1\%$)\\
  \hline
Strong & 144 ($15\%$) & 34 ($45\%$) & 127 ($19\%$) & 17 ($5\%$)\\
Weak &  100 ($10\%$) & 2 ($3\%$) & 61 ($9\%$) & 39 ($13\%$)\\
\hline
\end{tabular}
\label{Table:Lx}
\end{table}
%

\subsection{Radio Luminosity}
\begin{figure}
\centering
 \resizebox{\columnwidth}{!}
 {\includegraphics{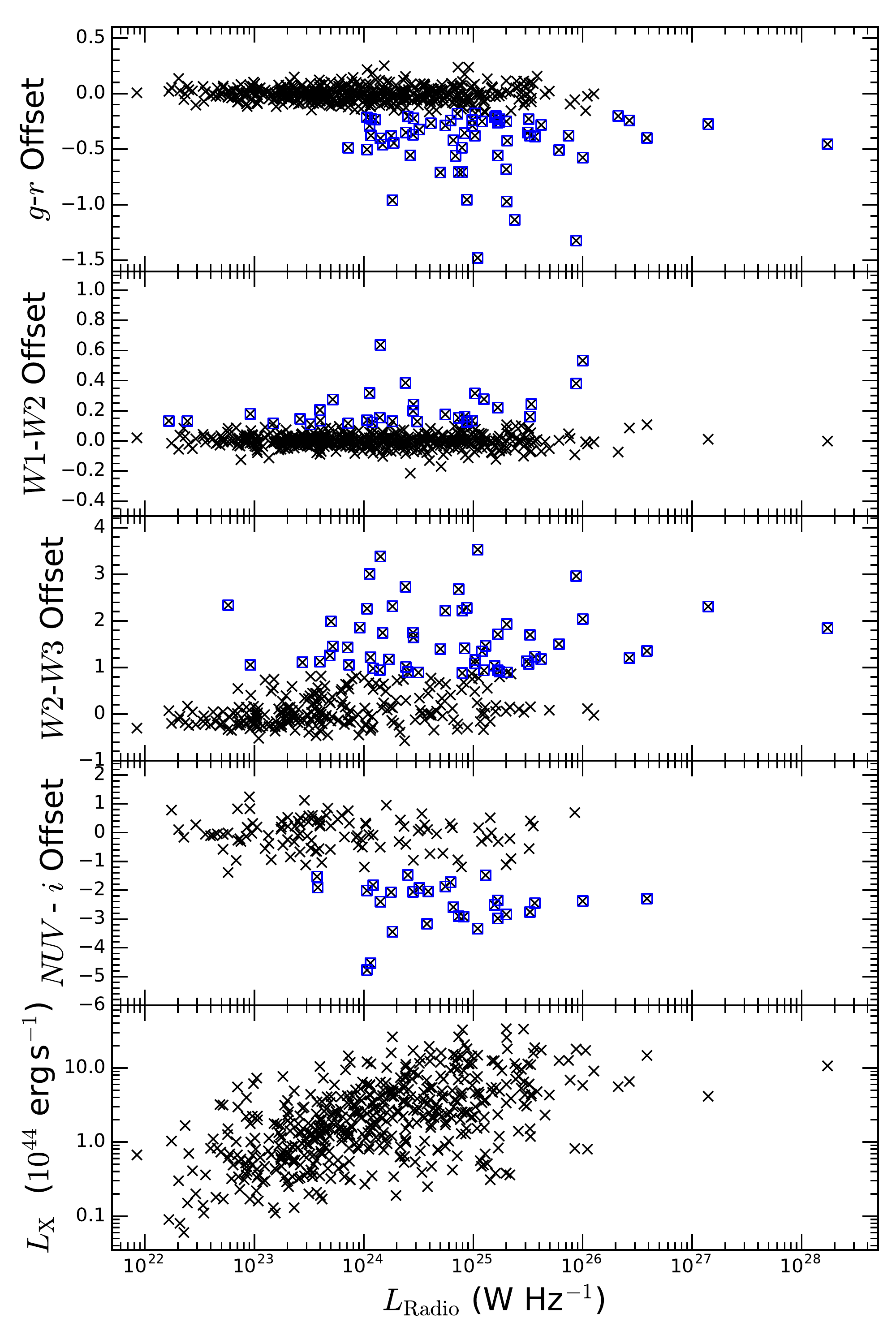}}
 \caption{The $1.4\,\mathrm{GHz}$ radio luminosity of the BCGs against the optical, Mid-IR and UV colour offsets (top four panels) and host cluster X-ray luminosity (bottom panel). The (blue) squares are indicative of offsets in excess of $2.5\,\sigma_p$.}
\label{Fig:Lr-Lx}
\end{figure}

\indent We find \RadioDet BCGs ($52\%$ of total sample) have a $1.4$ GHz radio detections within $45\arcsec$ in NVSS. In the colour offset against radio luminosity plots shown in Fig. \ref{Fig:Lr-Lx} there appears to be a connection between the radio emission and colour offsets. The apparent trends in the top four panels are likely to be a reflection of the relation between X-ray luminosity and radio power \citep{Hogan+15}, shown in the bottom panel, rather than suggesting that the star formation is induced or triggered by the radio activity in the BCG. This does however support the idea of a feedback cycle between radiative cooling induced star formation and AGN activity. Whereby relic radio emission still exists from previous stages of AGN activity in these galaxies.  
\\
\\

\begin{table}
 \centering
 \caption{Same as Table \ref{tab:fract}, but for the ``X-ray luminosity complete'' subsample of clusters with \Lx$\geq 1\times10^{44}\,\mathrm{erg\,s}^{-1}$ and $z\leq 0.2$.}
 \begin{tabular}{lcccccc}
  \hline
  \hline
& BCGs & Opt. & MIR & UV & Combined & Radio\\
& (\#) & (\%) & (\%) & (\%) & (\%) & (\%)\\ 
  \hline
All & 359 & 5 & 11 & 4 & 14  & 53\\
SL & 61 & 23 & 31 & 16 & 43 & 90 \\
WL & 39 & 5 & 10 & 5 & 15 & 67 \\
NL & 177 & 0 & 5 & 1 & 6  & 42\\
\hline
\label{Table:completefractions}
\end{tabular}
\end{table}

\begin{table}
 \centering
 \caption{Same as Table \ref{tab:fract}, but for the subsample of clusters with \Lx$\geq 1\times10^{44}\,\mathrm{erg\,s}^{-1}$ and $z\leq 0.1$ to match the sample selection in \protect\cite{Fraser-McKelvie+15}.}
 \begin{tabular}{lcccccc}
  \hline
  \hline
      & BCGs & Opt. & MIR & UV & Combined & Radio\\
    & (\#) & (\%) & (\%) & (\%) & (\%) & (\%)\\ 
  \hline
All & 95 & 3 & 8 & 2 & 11  & 61\\
SL & 17 & 18 & 29 & 6 & 35 & 88 \\
WL & 14 & 0 & 7 & 0 & 7 & 86 \\
NL & 52 & 0 & 0 & 2 & 2  & 48\\
\hline
\label{Table:FM+15Fractions}
\end{tabular}
\end{table}

\section{Conclusions}\label{sect:conclude}
We investigated the PS1 optical, \wise Mid-IR and \galex UV photometry for Brightest Cluster Galaxies in \TotalSample \textit{ROSAT} All Sky Survey X-ray selected clusters. The sample consists of clusters in the redshift range, $0.03<z<0.5$ with an X-ray luminosity range of $5\times 10^{43}\,\mathrm{erg\,s}^{-1}<L_X <5 \times10^{45}\,\mathrm{erg\,s}^{-1}$. The principal aim was to search for signs of star formation and/or AGN activity in BCGs by looking for photometric colours which deviate from that expected for a passive BCG. The main results are summarised as follows: 
\begin{itemize}
\item At least $14\%$ of the BCGs in our full sample show a significant colour deviation from passivity in at least one colour. In the optical we find $8\%$ are significantly blue in $PS1$ \grb. In the Mid-IR we find at least $9\%$ show a significant colour offset in either W1--W2 or W2--W3, with $5\%$ showing enhanced \wb2 emission, and at least $8\%$ showing enhanced \wb3 emission. In the NUV we find at least $4\%$ of BCGs show an enhanced NUV emission, shown as a significant NUV-\textit{i} colour offset. We stress that due to incompleteness these fractions represent lower limits for the NUV and \wb3. We interpret these colour deviations as a result of star formation and/or AGN activity within the BCG. Details regarding each of these active BCG candidates are provided in Table \ref{tab:active}.
\item We find that across all wavelengths the presence of optical emission lines and atypical photometric colours are strongly associated. This reinforces our interpretation of these BCGs being ``active''. We find that the majority of BCGs are non-line emitting and do not exhibit signs of enhanced emission, consistent with passivity.
\item We find that BCG activity has a strong association with the host cluster X-ray luminosity. Specifically, the fraction of peculiarly coloured BCGs is much higher in high \Lx clusters. This has important implications for the apparent prevalence of active BCGs and the selection function of a cluster sample. Additionally, there is an active identification bias against BCGs which host AGN being included in a cluster sample given the presumption that the AGN will be the majority contributor to the X-ray emission, when in fact the cluster emission alone may be still be sufficient to reach the selection flux limit.
\item We find that significant colour offsets in one wavelength regime is often associated with a significant offset in another, but not always. In the former case this illustrates the common origin of these colours at different wavelengths, but in the latter shows the benefit of a multi-wavelength analysis, in which we can trace activity over complimentary time-scales and degrees of obscuration.  
\item We find that the BCG luminosity, in both the $i$ and \wb1 bands, correlates with the X-ray luminosity of the host cluster. Specifically the most luminous, and hence most massive, BCGs tend to reside in the more X-ray luminous, and hence massive,  clusters.
\item The optical colours of the cluster red sequence are remarkably tightly constrained given the range in \Lx, indicative of the homogeneity in cluster galaxies across mass and redshift. For the majority of BCGs, this is also the case. So where BCGs are passively evolving, they do so in a similar fashion to one another. 
\item We have shown we are able to measure photometric excesses in active BCGs up to at least $z=0.5$, with potential to go to higher redshift. Hence, the same analysis could be applied to future, larger cluster samples. This creates opportunity with the new generation of X-ray telescopes, such as eROSITA, due to make observations in the near future, which will fill in the \Lx--$z$ parameter space. Combined with large photometric surveys such as the LSST one could identify a large number of ``active'' BCG candidates, which would be vital in really constraining the prevalence of activity in BCGs and better understanding the role of feedback in these systems.
\end{itemize}

\section*{Acknowledgements}
We thank the anonymous referee for their contribution in improving this manuscript. TSG received financial support from the Science and Technology Facilities Council (STFC) grant ST/K501979/1. ACE acknowledges support from STFC grant ST/I001573/1. JPS gratefully acknowledges a Hintze Research Fellowship. \\ \indent The Pan-STARRS1 Surveys (PS1) have been made possible through contributions by the Institute for Astronomy, the University of Hawaii, the Pan-STARRS Project Office, the Max-Planck Society and its participating institutes, the Max Planck Institute for Astronomy, Heidelberg and the Max Planck Institute for Extraterrestrial Physics, Garching, The Johns Hopkins University, Durham University, the University of Edinburgh, the Queen's University Belfast, the Harvard-Smithsonian Center for Astrophysics, the Las Cumbres Observatory Global Telescope Network Incorporated, the National Central University of Taiwan, the Space Telescope Science Institute, and the National Aeronautics and Space Administration under Grant No. NNX08AR22G issued through the Planetary Science Division of the NASA Science Mission Directorate, the National Science Foundation Grant No. AST-1238877, the University of Maryland, Eotvos Lorand University (ELTE), and the Los Alamos National Laboratory.  \\ \indent AllWISE makes use of data from WISE, which is a joint project of the University of California, Los Angeles, and the Jet Propulsion Laboratory/California Institute of Technology, and NEOWISE, which is a project of the Jet Propulsion Laboratory/California Institute of Technology. WISE and NEOWISE are funded by the National Aeronautics and Space Administration. \\ \indent This research has made use of the NASA/IPAC Extragalactic Database (NED) which is operated by the Jet Propulsion Laboratory, California Institute of Technology, under contract with the National Aeronautics and Space Administration.
 
\bibliographystyle{mnras}
\bibliography{PhotometricCensus} 

\appendix
\section{Table of ``active'' BCG candidates}
\begin{table*}
\caption{A table of those BCGs measured to be significantly offset in one colour or more (i.e. a colour offset in excess of $2.5\,\sigma_p$). The table key is as follows: (0) Cluster X-ray luminosity, in 0.1--2.4 keV band, in units of $10^{44} \mathrm{erg\,s}^{-1}$, (1) indicates whether the BCG has strong (S), weak (W) or no (N) optical emission lines or if no spectra available (-) (2) a PS1 \grb colour offset $<-0.18$ mags, (3) a \wise \wc offset $>0.11$ mags, (4) a \wise \wcol colour offset $>0.87$ mags and (5) a \galex NUV-$i$ colour offset $<-1.43$ mags. (a) The symbols in these columns are as follows: If a number is present this is the colour offset value and indicates a significant deviation from passivity, `$\times$' indicates a lack of a significant offset, `-' indicates this BCG not observed, and `..' indicates a poor $S/N$.} 
\label{tab:active}
\begin{tabular}{lcccrccccc}
\hline
\hline
\centering Cluster ID & BCG R.A. & BCG DEC & Redshift & \Lx & Lines & \grb & \wc & \wcol & NUV-$i$ \\
\centering & (J2000) & (J2000) &&(0) & (1) & (2a) & (3a) & (4a) & (5a) \\
\hline
RXJ0008.0-1151 & 00:07:57.0 & $-11$:51:44  & 0.21 & 1.48 & N & $\times$ & $\times$ & .. & -1.90 \\ 
RXJ0010.9+2909 & 00:10:53.4 & $+29$:09:39  & 0.33 & 4.29 & W & $\times$ & $\times$ & 1.21 & -4.53 \\ 
A11 & 00:12:33.8 & $-16$:28:07  & 0.15 & 0.88 & S & -0.29 & 0.18 & 2.22 & -1.87 \\ 
A22 & 00:20:43.1 & $-25$:42:28  & 0.14 & 3.46 & N & $\times$ & 0.12 & .. & - \\ 
MACSJ0025.4-1222 & 00:25:33.0 & $-12$:23:16  & 0.48 & 11.45 & - & -1.13 & $\times$ & .. & - \\ 
RXJ0045.9-1723 & 00:45:54.8 & $-17$:23:29  & 0.06 & 0.15 & W & $\times$ & 0.13 & $\times$ & - \\ 
S84 & 00:49:22.8 & $-29$:31:12  & 0.11 & 1.77 & S & $\times$ & 0.18 & 1.06 & .. \\ 
Z348 & 01:06:49.4 & $+01$:03:22  & 0.25 & 7.38 & S & -0.68 & 0.20 & 2.54 & -3.60 \\ 
A145 & 01:06:53.3 & $-02$:28:56  & 0.19 & 1.30 & N & $\times$ & $\times$ & 1.00 & - \\ 
MACSJ0111.5+0855 & 01:11:31.5 & $+08$:55:41  & 0.48 & 15.04 & N & -0.33 & $\times$ & .. & - \\ 
MACSJ0150.3-1005 & 01:50:21.3 & $-10$:05:30  & 0.36 & 6.61 & S & -0.42 & 0.18 & 2.40 & - \\ 
MACSJ0159.8-0849 & 01:59:49.3 & $-08$:49:58  & 0.40 & 18.06 & S & -0.43 & $\times$ & 0.89 & - \\ 
A291 & 02:01:43.1 & $-02$:11:48  & 0.20 & 5.97 & S & -0.21 & $\times$ & .. & -2.01 \\ 
A368 & 02:37:27.7 & $-26$:30:28  & 0.22 & 4.23 & S & $\times$ & $\times$ & 1.43 & - \\ 
RXJ0238.8-1258 & 02:38:49.3 & $-12$:58:49  & 0.20 & 2.53 & N & $\times$ & $\times$ & 0.88 & $\times$ \\ 
MACSJ0242.5-2132 & 02:42:35.9 & $-21$:32:25  & 0.31 & 14.74 & S & -0.40 & $\times$ & 1.35 & -2.30 \\ 
A383 & 02:48:03.3 & $-03$:31:44  & 0.19 & 5.09 & S & $\times$ & $\times$ & $\times$ & -2.04 \\ 
A3088 & 03:07:02.2 & $-28$:39:56  & 0.25 & 7.48 & S & -0.27 & $\times$ & $\times$ & - \\ 
MACSJ0326.8-0043 & 03:26:49.9 & $-00$:43:51  & 0.45 & 14.29 & S & -0.49 & $\times$ & .. & - \\ 
MACSJ0329.6-0211 & 03:29:41.6 & $-02$:11:46  & 0.45 & 11.85 & S & -0.71 & $\times$ & 1.39 & - \\ 
RXJ0331.1-2100 & 03:31:05.9 & $-21$:00:32  & 0.19 & 5.48 & S & -0.20 & $\times$ & $\times$ & - \\ 
RXJ0353.0+1941 & 03:52:59.0 & $+19$:40:59  & 0.11 & 2.47 & S & $\times$ & 0.28 & 1.46 & - \\ 
MACSJ0404.6+1109 & 04:04:33.7 & $+11$:07:53  & 0.36 & 11.35 & - & -0.29 & 0.14 & .. & - \\ 
A483 & 04:15:57.6 & $-11$:32:53  & 0.28 & 5.27 & - & $\times$ & $\times$ & 0.97 & - \\ 
MACSJ0417.5-1154 & 04:17:34.7 & $-11$:54:32  & 0.44 & 33.83 & S & -0.68 & $\times$ & .. & -2.84 \\ 
MACSJ0429.6-0253 & 04:29:36.0 & $-02$:53:06  & 0.40 & 12.64 & S & -0.38 & $\times$ & .. & - \\ 
RXJ0437.1+0043 & 04:37:09.6 & $+00$:43:51  & 0.28 & 8.68 & N & -0.22 & $\times$ & .. & - \\ 
RXJ0439.0+0520 & 04:39:02.3 & $+05$:20:42  & 0.21 & 5.83 & S & -0.24 & $\times$ & $\times$ & - \\ 
RXJ0448.2+0952 & 04:48:12.8 & $+09$:53:01  & 0.15 & 2.90 & - & -0.23 & $\times$ & .. & - \\ 
A521 & 04:54:06.9 & $-10$:13:24  & 0.25 & 9.26 & N & $\times$ & $\times$ & 1.56 & - \\ 
RXJ0505.2-0217 & 05:05:16.7 & $-02$:19:24  & 0.23 & 3.05 & - & $\times$ & $\times$ & 1.13 & - \\ 
RXJ0524.3+0329 & 05:24:19.1 & $+03$:29:54  & 0.15 & 2.43 & - & $\times$ & 0.16 & .. & - \\ 
RXJ0611.0-2735 & 06:11:01.2 & $-27$:35:33  & 0.04 & 0.15 & S & $\times$ & 0.27 & 1.27 & -2.31 \\ 
Z1121 & 06:31:22.7 & $+25$:01:06  & 0.08 & 2.38 & W & - & 0.15 & 0.99 & - \\ 
CIZAJ0640.1-1253 & 06:40:07.2 & $-12$:53:14  & 0.14 & 7.18 & - & -0.38 & 0.32 & 1.09 & - \\ 
CIZAJ0710.4+2240 & 07:10:23.8 & $+22$:40:02  & 0.29 & 11.30 & - & -0.35 & $\times$ & .. & .. \\ 
CIZAJ0719.5+0043 & 07:19:36.0 & $+00$:42:33  & 0.22 & 9.05 & - & -0.19 & $\times$ & .. & - \\ 
PKS745-191 & 07:47:31.4 & $-19$:17:39  & 0.10 & 12.50 & S & -0.51 & $\times$ & 1.50 & - \\ 
RXJ0815.5-0308 & 08:15:27.8 & $-03$:08:27  & 0.20 & 5.58 & - & -0.20 & $\times$ & .. & - \\ 
RXJ0821.0+0751 & 08:21:02.3 & $+07$:51:47  & 0.11 & 1.34 & S & $\times$ & $\times$ & 2.46 & -2.16 \\ 
RXJ0834.9+5534 & 08:34:55.0 & $+55$:34:20  & 0.24 & 4.14 & S & -0.28 & $\times$ & 2.31 & $\times$ \\ 
Z1883 & 08:42:55.9 & $+29$:27:26  & 0.19 & 3.37 & S & -0.20 & $\times$ & 0.90 & -1.47 \\ 
Z2089 & 09:00:37.0 & $+20$:53:40  & 0.23 & 6.46 & S & -0.40 & 0.64 & 3.38 & -2.40 \\ 
MACSJ0913.7+4056 & 09:13:45.6 & $+40$:56:27  & 0.44 & 14.87 & S & -1.48 & 1.77 & 3.52 & -3.34 \\ 
Hyd-A & 09:18:05.7 & $-12$:05:43  & 0.05 & 6.56 & S & -0.24 & $\times$ & 1.21 & - \\ 
A854 & 09:42:01.1 & $+08$:58:19  & 0.21 & 3.40 & N & $\times$ & $\times$ & 1.26 & - \\ 
RBS797 & 09:47:12.7 & $+76$:23:13  & 0.34 & 20.69 & S & -0.36 & 0.16 & 1.41 & -2.88 \\ 
RXJ1000.4+4409 & 10:00:31.0 & $+44$:08:43  & 0.15 & 2.25 & W & $\times$ & 0.14 & .. & $\times$ \\ 
A910 & 10:03:00.2 & $+67$:07:56  & 0.21 & 3.47 & - & $\times$ & $\times$ & $\times$ & -1.53 \\ 
A926 & 10:06:40.2 & $+21$:40:14  & 0.18 & 1.70 & S & $\times$ & 0.20 & 1.12 & - \\ 
MS1006.0+1202 & 10:08:47.7 & $+11$:47:38  & 0.26 & 3.87 & W & $\times$ & $\times$ & 1.85 & - \\ 
Z3146 & 10:23:39.6 & $+04$:11:10  & 0.29 & 26.29 & S & -0.96 & 0.13 & 2.31 & -3.45 \\ 
A3444 & 10:23:50.2 & $-27$:15:23  & 0.25 & 15.96 & S & -0.38 & $\times$ & $\times$ & -2.07 \\ 
A1068 & 10:40:44.4 & $+39$:57:11  & 0.14 & 5.74 & S & -0.29 & 0.32 & 3.01 & - \\ 
A1084 & 10:44:32.9 & $-07$:04:07  & 0.13 & 4.10 & S & $\times$ & $\times$ & 0.95 & $\times$ \\ 
\hline
\end{tabular}
\end{table*}
\begin{table*}
\contcaption{A table continued from the previous one.}
\label{tab:continued}
\begin{tabular}{lcccrccccc}
\hline
\hline
\centering Cluster ID & BCG R.A. & BCG DEC & Redshift & \Lx & Lines & \grb & \wc & \wcol & NUV-$i$ \\
\hline
A1126 & 10:53:50.2 & $+16$:51:02  & 0.09 & 1.19 & - & $\times$ & $\times$ & 0.98 & $\times$ \\ 
A1211 & 11:14:50.3 & $-12$:13:50  & 0.19 & 3.21 & N & $\times$ & $\times$ & 1.05 & $\times$ \\ 
Z3959 & 11:15:51.8 & $+01$:29:54  & 0.35 & 15.29 & S & -0.42 & $\times$ & .. & -2.59 \\ 
RXJ1124.5+4351 & 11:24:29.7 & $+43$:51:25  & 0.37 & 8.02 & S & -0.34 & $\times$ & .. & - \\ 
RXJ1128.1+7529 & 11:28:09.9 & $+75$:29:35  & 0.17 & 0.93 & - & $\times$ & 0.16 & .. & - \\ 
MACSJ1141.6-1905 & 11:41:40.8 & $-19$:05:15  & 0.30 & 3.84 & - & $\times$ & 0.13 & .. & - \\ 
A1451 & 12:03:16.8 & $-21$:32:55  & 0.20 & 5.00 & N & $\times$ & $\times$ & 1.17 & - \\ 
MACSJ1218.4+4012 & 12:18:28.6 & $+40$:12:38  & 0.30 & 5.75 & - & -0.20 & $\times$ & .. & - \\ 
RXJ1259.1+4129 & 12:59:08.6 & $+41$:29:37  & 0.28 & 2.99 & W & -0.18 & $\times$ & .. & - \\ 
RXJ1301.0-2312 & 13:00:58.5 & $-23$:12:15  & 0.13 & 1.51 & S & -0.38 & 0.16 & 1.70 & -2.76 \\ 
A1664 & 13:03:42.6 & $-24$:14:42  & 0.13 & 4.24 & S & -0.46 & $\times$ & 1.75 & -3.12 \\ 
A1704** & 13:14:24.6 & $+64$:34:31  & 0.22 & 4.35 & N & -0.22 & $\times$ & 1.10 & -2.09 \\ 
A1750b & 13:31:11.0 & $-01$:43:38  & 0.08 & 0.61 & - & $\times$ & $\times$ & 2.34 & $\times$ \\ 
RXJ1336.0-0331 & 13:35:60.0 & $-03$:31:28  & 0.17 & 1.61 & W & $\times$ & $\times$ & $\times$ & -1.91 \\ 
MACSJ1347.5-1144 & 13:47:30.7 & $-11$:45:09  & 0.45 & 49.41 & S & -0.72 & $\times$ & .. & - \\ 
A1795 & 13:48:52.5 & $+26$:35:34  & 0.06 & 9.19 & S & $\times$ & $\times$ & $\times$ & -2.92 \\ 
RBS1322 & 13:50:22.1 & $+09$:40:10  & 0.13 & 3.67 & S & $\times$ & $\times$ & 1.46 & -1.48 \\ 
MACSJ1354.6+7715 & 13:54:43.0 & $+77$:15:16  & 0.40 & 6.60 & S & -0.24 & $\times$ & 1.69 & - \\ 
RXJ1359.3+7447 & 13:59:16.9 & $+74$:46:42  & 0.20 & 1.68 & W & -0.23 & $\times$ & $\times$ & - \\ 
RXJ1359.9+1414 & 13:59:57.3 & $+14$:14:17  & 0.21 & 2.67 & W & $\times$ & $\times$ & 1.01 & .. \\ 
A1835 & 14:01:02.2 & $+02$:52:42  & 0.25 & 26.31 & S & -0.71 & 0.15 & 2.67 & -2.90 \\ 
RXJ1401.3+2501 & 14:01:17.8 & $+25$:01:50  & 0.41 & 6.30 & - & -0.18 & $\times$ & .. & - \\ 
MACSJ1411.3+5212 & 14:11:20.4 & $+52$:12:09  & 0.46 & 10.67 & S & -0.45 & $\times$ & 1.84 & - \\ 
A1910 & 14:24:24.4 & $+25$:14:28  & 0.23 & 2.41 & S & -0.50 & 0.14 & 2.26 & -4.77 \\ 
A1918 & 14:25:22.5 & $+63$:11:53  & 0.14 & 2.45 & N & $\times$ & $\times$ & $\times$ & -1.58 \\ 
RXJ1427.2+4407 & 14:27:16.1 & $+44$:07:30  & 0.49 & 17.39 & S & -0.28 & $\times$ & 1.17 & - \\ 
RXJ1434.7+1721 & 14:34:42.9 & $+17$:21:57  & 0.04 & 0.09 & N & $\times$ & 0.13 & $\times$ & - \\ 
RXJ1447.4+0827 & 14:47:26.0 & $+08$:28:25  & 0.38 & 25.77 & S & -0.97 & 0.11 & 1.92 & - \\ 
Z7160 & 14:57:15.1 & $+22$:20:34  & 0.26 & 11.54 & S & -0.32 & $\times$ & $\times$ & -1.92 \\ 
RXJ1459.1-0842 & 14:59:05.2 & $-08$:42:36  & 0.10 & 1.15 & N & $\times$ & 0.12 & $\times$ & - \\ 
S780 & 14:59:28.8 & $-18$:10:45  & 0.24 & 11.07 & S & -0.56 & $\times$ & 1.71 & -2.98 \\ 
RXJ1504.1-0248 & 15:04:07.5 & $-02$:48:16  & 0.22 & 32.73 & S & -0.71 & $\times$ & 2.22 & -4.72 \\ 
RXJ1512.8-0127 & 15:12:52.6 & $-01$:28:26  & 0.12 & 0.78 & N & $\times$ & $\times$ & 1.01 & - \\ 
A2055 & 15:18:45.7 & $+06$:13:56  & 0.10 & 3.62 & W & $\times$ & 0.28 & 0.94 & - \\ 
MACSJ1532.8+3021 & 15:32:53.8 & $+30$:20:59  & 0.34 & 18.10 & S & -0.95 & 0.13 & 2.28 & - \\ 
A2104 & 15:40:07.9 & $-03$:18:15  & 0.15 & 5.26 & N & $\times$ & 0.20 & $\times$ & - \\ 
MACSJ1551.9-0207 & 15:51:58.6 & $-02$:07:50  & 0.30 & 5.47 & - & -0.44 & $\times$ & .. & - \\ 
A2146 & 15:56:14.2 & $+66$:20:52  & 0.23 & 7.82 & S & $\times$ & 0.38 & 2.73 & .. \\ 
PKS1555-140 & 15:58:21.9 & $-14$:09:58  & 0.10 & 4.57 & S & $\times$ & $\times$ & 1.16 & - \\ 
RXJ1600.0-0354 & 16:00:02.5 & $-03$:54:35  & 0.27 & 4.39 & - & -0.24 & $\times$ & .. & - \\ 
A2147b & 16:03:38.1 & $+15$:54:02  & 0.11 & 2.69 & S & $\times$ & 0.24 & 1.65 & $\times$ \\ 
RXJ1614.3+5442 & 16:14:15.4 & $+54$:43:28  & 0.33 & 3.53 & N & -0.18 & $\times$ & .. & - \\ 
MACSJ1621.3+3810 & 16:21:24.7 & $+38$:10:08  & 0.47 & 12.38 & S & -0.56 & $\times$ & .. & - \\ 
A2204 & 16:32:47.0 & $+05$:34:31  & 0.15 & 14.97 & S & -0.27 & $\times$ & $\times$ & - \\ 
RBS1634 & 17:17:07.0 & $+29$:31:21  & 0.28 & 14.57 & S & -0.49 & $\times$ & .. & - \\ 
Z8193 & 17:17:19.2 & $+42$:26:57  & 0.18 & 3.39 & S & -0.25 & $\times$ & 1.35 & - \\ 
Z8197 & 17:18:11.9 & $+56$:39:56  & 0.11 & 2.27 & S & $\times$ & $\times$ & $\times$ & -1.82 \\ 
RXJ1720.2+2637 & 17:20:10.1 & $+26$:37:32  & 0.16 & 9.29 & S & -0.24 & $\times$ & $\times$ & -1.72 \\ 
A2262 & 17:23:21.6 & $+23$:50:39  & 0.23 & 4.10 & - & $\times$ & $\times$ & 1.98 & - \\ 
A2270 & 17:27:23.5 & $+55$:10:53  & 0.24 & 2.80 & S & -0.39 & 0.30 & 2.36 & - \\ 
Z8276 & 17:44:14.5 & $+32$:59:29  & 0.08 & 3.29 & S & $\times$ & $\times$ & 0.98 & - \\ 
CIZAJ1804.1+0042 & 18:04:08.9 & $+00$:42:22  & 0.09 & 2.89 & - & $\times$ & $\times$ & 0.89 & - \\ 
RXJ1817.8+6824 & 18:17:44.5 & $+68$:24:25  & 0.29 & 3.38 & - & -0.44 & $\times$ & $\times$ & - \\ 
RGBJ1832+688 & 18:32:35.8 & $+68$:48:05  & 0.20 & 4.86 & S & -0.25 & $\times$ & 0.91 & .. \\ 
CIZAJ1904.2+3627 & 19:04:11.9 & $+36$:26:59  & 0.08 & 0.90 & - & $\times$ & 0.16 & $\times$ & - \\ 
CIZAJ1917.6-1315 & 19:17:36.2 & $-13$:15:11  & 0.18 & 3.25 & - & $\times$ & $\times$ & .. & -3.17 \\ 
MACSJ1931.8-2634 & 19:31:49.7 & $-26$:34:32  & 0.35 & 17.98 & S & -1.32 & 0.38 & 2.96 & - \\ 
Cyg-A & 19:59:28.3 & $+40$:44:02  & 0.06 & 4.35 & S & -0.49 & 0.96 & 3.69 & - \\ 
MRC2011-246 & 20:14:51.8 & $-24$:30:23  & 0.16 & 12.56 & S & -0.21 & $\times$ & 1.03 & -2.51 \\ 
\hline
\end{tabular}
{~~~~~~~~~~~~ **The BCG of A1704 has an SDSS spectra on its core, which shows no emission lines. But in the imaging there is clearly an optically blue component within the BCG and just North of its core, which does not have spectroscopy. We detect this in our photometry and suspect this is associated with activity in the BCG, but projection cannot be ruled out.}
\end{table*}
\begin{table*}
\contcaption{A table continued from the previous one.}
\label{tab:continued}
\begin{tabular}{lcccrccccc}
\hline
\hline
\centering Cluster ID & BCG R.A. & BCG DEC & Redshift & \Lx & Lines & \grb & \wc & \wcol & NUV-$i$ \\
\hline
RXJ2020.3-2225 & 20:20:22.6 & $-22$:25:32  & 0.29 & 5.58 & - & $\times$ & $\times$ & 1.88 & - \\ 
RXJ2043.2-2144 & 20:43:14.6 & $-21$:44:34  & 0.20 & 4.46 & S & $\times$ & 0.24 & .. & - \\ 
RXJ2100.0-2426 & 20:59:55.6 & $-24$:25:45  & 0.08 & 0.28 & - & $\times$ & 0.22 & 1.89 & -1.60 \\ 
RXJ2125.4+1742 & 21:25:22.0 & $+17$:43:05  & 0.22 & 3.37 & - & -0.51 & $\times$ & 0.96 & - \\ 
MACSJ2134.6-2706 & 21:34:36.0 & $-27$:05:55  & 0.36 & 6.01 & - & -0.25 & $\times$ & .. & - \\ 
RGBJ2138+359 & 21:38:21.1 & $+35$:58:23  & 0.11 & 1.04 & - & $\times$ & 0.33 & $\times$ & - \\ 
MS2137.3-2353 & 21:40:15.2 & $-23$:39:40  & 0.31 & 11.36 & S & -0.38 & $\times$ & .. & - \\ 
MACSJ2149.3+0951 & 21:49:19.6 & $+09$:51:37  & 0.38 & 8.55 & - & -0.28 & $\times$ & .. & - \\ 
A2390 & 21:53:37.0 & $+17$:41:42  & 0.23 & 18.77 & S & -0.39 & $\times$ & 1.23 & -2.45 \\ 
RXJ2213.1-2754 & 22:13:05.9 & $-27$:54:20  & 0.03 & 0.13 & S & $\times$ & 0.12 & $\times$ & - \\ 
A2442 & 22:25:51.2 & $-06$:35:34  & 0.09 & 1.32 & N & $\times$ & $\times$ & .. & -1.46 \\ 
MACSJ2229.7-2755 & 22:29:45.2 & $-27$:55:35  & 0.32 & 11.19 & S & -0.35 & $\times$ & 1.01 & - \\ 
MACSJ2243.3-0935 & 22:43:20.7 & $-09$:35:18  & 0.45 & 21.05 & N & -0.20 & $\times$ & .. & - \\ 
CIZAJ2302.6+7136 & 23:02:38.6 & $+71$:36:25  & 0.14 & 2.87 & - & $\times$ & $\times$ & 1.11 & - \\ 
RXJ2311.3-0946 & 23:11:18.9 & $-09$:46:22  & 0.49 & 9.47 & S & -0.55 & $\times$ & .. & - \\ 
RXJ2320.9-0433 & 23:20:54.2 & $-04$:34:02  & 0.19 & 2.42 & N & $\times$ & $\times$ & 0.88 & - \\ 
A2597 & 23:25:19.7 & $-12$:07:26  & 0.09 & 5.99 & S & -0.23 & $\times$ & 1.08 & .. \\ 
RXJ2326.3-2406 & 23:26:14.2 & $-24$:06:30  & 0.06 & 0.19 & N & $\times$ & 0.14 & .. & - \\ 
A2627 & 23:36:42.1 & $+23$:55:29  & 0.12 & 2.35 & W & -0.26 & 0.22 & 0.94 & -2.35 \\ 
PKS2338+000 & 23:41:07.0 & $+00$:18:33  & 0.28 & 5.81 & S & -0.58 & 0.53 & 2.04 & -2.38 \\ 
A2667 & 23:51:39.4 & $-26$:05:02  & 0.23 & 17.13 & S & -0.37 & 0.20 & 1.74 & -2.06 \\ 
RXJ2355.4-1027 & 23:55:25.6 & $-10$:27:22  & 0.29 & 3.04 & N & $\times$ & 0.13 & .. & - \\ 
\hline
\end{tabular}
\end{table*}

\bsp	
\label{lastpage}
\end{document}